\documentclass[lettersize,journal]{IEEEtran}

\usepackage[english]{babel}
\usepackage[noadjust]{cite}
\usepackage{algorithm}
\usepackage{algorithmic}
\usepackage{etoolbox}
\usepackage{optidef}
\usepackage{comment}
\usepackage{graphicx}
\usepackage{url}

\usepackage{multirow}
\usepackage{tabularx}
\usepackage{graphicx}
\usepackage{svg}
\usepackage{optidef}
\usepackage{amsmath}
\usepackage{bm}
\usepackage{amssymb}
\usepackage{xcolor}
\usepackage{caption}
\usepackage{subcaption}
\usepackage{lipsum}
\usepackage{amsthm}
\usepackage{siunitx}
\usepackage[acronym]{glossaries}
\usepackage{graphicx}
\usepackage{subcaption}

\usepackage{booktabs}

\usepackage{titlesec}
\titlespacing\section{0pt}{*0.8}{*0.8}
\titlespacing\subsection{0pt}{*0.8}{*0.8}
\titlespacing\subsubsection{0pt}{*0.7}{*0.7}

\makeglossaries

\newacronym{psr}{PSR}{power split ratio}
\newacronym{isac}{ISAC}{integrated sensing and communications}
\newacronym{mimo}{MIMO}{Multiple-input multiple-output}
\newacronym{iv}{IV}{instrumental variable}
\newacronym{nsp}{NSP}{null-space projection}
\newacronym{dof}{DoF}{degree of freedom}
\newacronym{sinr}{SINR}{signal-to-interference-plus-noise ratio}
\newacronym{sdma}{SDMA}{spatial division multiplexing}
\newacronym{csi}{CSI}{channel state information}
\newacronym{mui}{MUI}{multi-user interference}
\newacronym{rcg}{RCG}{Riemannian conjugate gradient}
\newacronym{crb}{CRB}{Cramer-Rao bound}
\newacronym{lmmse}{LMMSE}{linear minimum mean square error}
\newacronym{isl}{ISL}{integrated sidelobe level}
\newacronym{papr}{PAPR}{peak-to-average power ratio}
\newacronym{swipt}{SWIPT}{simultaneous information and power transfer}
\newacronym{qos}{QoS}{quality-of-service}
\newacronym{sar}{SAR}{synthetic aperture radar}
\newacronym{svd}{SVD}{singular value decomposition}
\newacronym{xr}{XR}{extended reality}
\newacronym{3gpp}{3GPP}{Third Generaion Partnership Project}
\newacronym{nr}{NR}{New Radio}
\newacronym{prs}{PRS}{Positioning Reference Signal}
\newacronym{ambc}{AmBC}{Ambient Backscatter Communication}
\newacronym{iot}{IoT}{Internet-of-Things}
\newacronym{snr}{SNR}{signal-to-noise ratio}
\newacronym{irs}{IRS}{intelligent reflecting surfaces}
\newacronym{ris}{RIS}{reflecting intelligent surfaces}
\newacronym{espar}{ESPAR}{electronically steerable parasitic array radiator}
\newacronym{hfss}{HFSS}{Ansys high frequency structure simulator}
\newacronym{mamp}{MAMP}{multi-active multi-parasitic }

\newacronym{xlmimo}{XL-MIMO}{large-scale MIMO}
\newacronym{umi}{UMi}{urban-micro cell}
\newacronym{uma}{UMa}{urban-macro cell}
\newacronym{los}{LOS}{line-of-sight}
\newacronym{nlos}{NLOS}{non line-of-sight}
\newacronym{quadriga}{QuaDRiGa}{quasi deterministic radio channel generator}
\newacronym{gevp}{GEVP}{generalized eigen value problem}
\newacronym{kkt}{KKT}{Karush--Kuhn--Tucke}
\newacronym{gla}{GLA}{genralized Lloyd algorithm}

\begin{document}

\title{Parasitic MIMO Beamforming for Multi-Active Multi-Parasitic Antenna Arrays with Binary Control}

\author{
Taejun Lee, Byunghyun Lee, Thomas E. Roth, and David J. Love
\thanks{Taejun Lee, Thomas E. Roth, and David J. Love are with the Elmore Family School of Electrical and Computer Engineering, Purdue University, West Lafayette, IN 47907 USA (e-mail:lee5198@purdue.edu; rothte@purdue.edu; djlove@purdue.edu).
}
\thanks{Byunghyun Lee is with Qualcomm Technologies, Inc.
5775 Morehouse Drive, San Diego, CA 92121 USA (e-mail: byunlee@qti.qualcomm.com).}
}



\maketitle

\begin{abstract}
In 6G, multiple-input multiple-output dimensions will continue to scale, yet the increased cost, power consumption, and hardware complexity associated with a growing number of RF chains limit practical deployment.
Parasitic antennas offer a promising alternative that can add spatial degrees of freedom and array gain without a proportional increase in RF chains.
From a communication perspective, prior work on parasitic antennas has primarily focused on adjusting continuous reactance values using varactors, but such varactor-based tuning has increased cost and complexity in the analog control and practical RF circuit design.
This paper proposes a \gls{mamp} antenna architecture with binary controllers, where each parasitic element operates in one of two discrete reactance states.
To validate the practicality of the system, we experimentally identify array geometries that best match the actual radiation patterns with those of the mathematical model through full-wave electromagnetic simulations with HFSS.
We express the induced current vector as a quadratic function of the binary state vector, and propose a pair of discrete reactance values that minimize the relative error of the proposed model while being implementable with off-the-shelf RF components.
With these results, we develop two transmit beamforming codebook designs based on the generalized Lloyd algorithm.
The first design exhaustively searches for all possible binary combinations to find the optimal solution, representing the theoretical upper limits of our framework.
The second design leverages eigenvalue perturbation to significantly reduce computational complexity, making it suitable for online adaptation.
Extensive simulations under various channel scenarios demonstrate that the proposed codebook designs enable \gls{mamp} with only few active antennas to achieve beamforming performance comparable to fully active antenna arrays with significantly more active antennas.
\end{abstract}

\begin{IEEEkeywords}
Multiple-Active Multiple-Parasitic (MAMP), ESPAR, binary control, limited feedback
\end{IEEEkeywords}

\IEEEpeerreviewmaketitle

\glsresetall

\section{Introduction}

\gls{mimo} has been developed as one of the most fundamental and significant technologies in wireless communications. Its importance persists in next-generation systems, including the 6th generation of wireless communication and beyond \cite{heath2025tri, brinton2025key}. By increasing the number of antennas, \gls{mimo} systems can achieve both diversity gain and multiplexing gain, which motivates the development of ever-larger antenna arrays ranging from massive \gls{mimo} to \gls{xlmimo} \cite{wang_tutorial_2024}. However, in conjunction with this trend, challenges related to energy and cost efficiency have become increasingly critical. As a result, a variety of energy and cost-efficient antenna array architectures and techniques have been proposed \cite{castellanos_embracing_2025}.

Among them, \gls{ris} have attracted substantial research interest since the early 2020s~\cite{wu_intelligent_2019, wu_towards_2020, gong_toward_2020}. However, \gls{ris} requires the deployment of reflective elements throughout the surrounding environment, which makes performance highly environment-dependent and raises concerns about its practical viability. Moreover, its passive nature presents significant challenges in terms of signaling and control, as it lacks active transmission capabilities, making real-time adaptation and feedback coordination difficult.
Several challenges related to the commercialization of \gls{ris} in mobile networks, such as interference characteristics and deployment constraints, are discussed in detail in \cite{astrom_ris_2024}. Due to these concerns, the \gls{3gpp} Release 19 no longer considers IRS as an official study item, but rather leaves it as a potential candidate for future specifications.

As an energy- and cost-efficient alternative, research interest in parasitic antenna arrays has been growing, as they require significantly fewer RF chains compared to the total number of antenna elements \cite{kalis_parasitic_2014}. Parasitic antenna arrays achieve beamforming control by leveraging mutual coupling between active elements, which are connected to RF chains, and surrounding parasitic elements.
To induce strong mutual coupling between antennas, these arrays allow for closer spacing between antenna elements than conventional arrays, which helps reduce the overall physical size of the antenna array. This makes them well-suited for compact devices where cost and efficiency are critical, and also offers advantages from an antenna deployment perspective.

Research on parasitic antenna arrays was first introduced in \cite{gyoda_design_2000} under the concept of \gls{espar}. In \gls{espar}, a single active antenna is surrounded by multiple parasitic elements whose mutual coupling and reactance are tuned to achieve beamforming. Transmit beamforming in single-active configurations has been studied in \cite{zhang_analog_2023}, while receive beamforming has been examined in \cite{bucheli_garcia_low-complexity_2020}. However, these works are inherently limited in scalability as they focus solely on single-active antenna cases. Extensions to multiple-active \gls{espar} configurations have also been explored in \cite{deshpande_beamforming_2025, kalis_parasitic_2014}.

Beyond beamforming, ESPAR-based approaches have also been investigated for spatial multiplexing \cite{han_single-rf_2023, han_mimo_2013, lee_spatial_2017, barousis_beamspace-domain_2011}. These techniques aim to identify orthogonal radiation patterns in the beamspace domain and assign symbols to each pattern. The works in \cite{han_mimo_2013, lee_spatial_2017, barousis_beamspace-domain_2011} propose orthogonalizing radiation patterns using the Gram–Schmidt process,  
while \cite{han_single-rf_2023} introduces a new method to construct an orthogonal basis set based on the theory of characteristic modes (TCM) \cite{han_characteristic_2021}. However, this requires determining appropriate reactance combinations for every transmitted symbol, and the number of required reactance states grows rapidly with the modulation order and the number of antennas. To mitigate this complexity, \cite{han_single-rf_2023} proposed a beamspace space-shift keying scheme as an alternative modulation method.

Despite these promising directions, a key drawback of existing \gls{espar} designs lies in their reliance on continuous reactance tuning using RF components such as varactors. Compared with switched parasitic antenna designs, this approach results in higher hardware complexity, implementation cost, and lower scalability~\cite{thiel_switched_2004}. Varactor-based designs require more complex and expensive control circuitry, regardless of whether for direct analog control or for digitized voltage control. 
Furthermore, maintaining accurate voltage-to-reactance characteristics becomes more difficult as the number of parasitic antennas increases, making large-scale implementations less practical.
Due to these limitations, many studies in the antenna community have investigated switched parasitic antenna designs using simpler RF components such as PIN diodes or MEMS switches~\cite{vaughan_switched_1999, thiel2002switched, petit_mems-switched_2006, kausar_energy_2016}. However, their primary focus has been on antenna design and RF implementation, whereas  beamforming and codebook design for switched \gls{mamp} in the communication society remain largely unexplored.

Unlike existing switched parasitic antenna studies, which primarily investigate radiation characteristics and beam steering~\cite{kalis_parasitic_2014}, this paper establishes the first mathematical communication framework for binary-switched MAMP antenna arrays. The proposed framework models the transmitted and received signals as explicit functions of the binary switching vector and derives a polynomial approximation for efficient beamforming codebook design.
Furthermore, much of the prior \gls{espar} literature relies on beamspace-domain analysis proposed in \cite{sayeed_deconstructing_2002}, where far-field radiation patterns play a central role. While such analysis is meaningful when antenna spacing exceeds half a wavelength, parasitic arrays typically involve much closer element placement. This can induce unexpected higher-order scattering, leading to discrepancies between theoretical predictions and actual radiation patterns. To address this, we perform detailed HFSS simulations to experimentally identify antenna parameter values that minimize the discrepancy between the mathematical model and the actual radiation behavior.

The primary contributions of this paper are summarized below:
\begin{itemize}
    \item We propose a \gls{mamp} antenna architecture controlled by on--off switches. Through \gls{hfss} simulations, we experimentally identify antenna configurations that best match with the mathematical model, thereby validating the proposed model.
    \item We formulate the on-off states of the parasitic antennas as a binary vector and derive the induced current vector as a quadratic function of the binary vector. Unlike conventional switched antenna models that focus on radiation pattern synthesis and beam steering, the derived current vector is interpreted as the transmit precoding vector, enabling communication-theoretic beamforming and codebook design.
    \item We extend the limited feedback framework widely adopted in 4G and 5G systems \cite{narula2002performance, love2004value} by developing an offline codebook design method based on the \gls{gla}, where each codeword is represented by a pair of a binary vector and an input voltage vector. Since the centroid condition finds the optimal solution within a given channel region through exhaustive search, it serves as a performance upper bound in the proposed model.
    \item We develop a low-complexity greedy codebook design, in which the on-state antennas are determined based on top scores computed using eigenvalue perturbation. Since the computational complexity of the proposed algorithm is even lower than that of the exhaustive search method, it enables online codebook design and real-time updates in response to user mobility or frequent changes in channel statistics.
\end{itemize}

The remainder of this paper is organized as follows. In Section \ref{section2}, we introduce the system and channel models in the beamspace domain and present the current vector for the \gls{mamp} architecture.
In Section \ref{section3}, we formulate the current vector as a quadratic function of the binary state vector, and identify practical antenna configurations and two-state reactance values through \gls{hfss}.
Section \ref{section4} proposes a codebook design scheme for transmit beamforming based on exhaustive search, while Section \ref{section5} introduces a low-complexity extension using eigenvalue perturbation.
In Section \ref{section6}, we evaluate the performance of the proposed codebooks in terms of beamforming gain. The conclusions are presented in Section \ref{section7}.

\textbf{Notation}:
Vectors and matrices are denoted by boldface lowercase and uppercase letters.
The operators $(\cdot)^T$, $(\cdot)^H$, and $(\cdot)^{-1}$ indicate transpose, transpose, and inverse operators, respectively.
The symbols $|\cdot|$ and $\Vert \cdot\Vert$ stand for the absolute and 2-norm operators.
$\operatorname{diag}(\cdot)$ forms a diagonal matrix from a given vector.
$\mathbb{E}[\cdot]$ denotes the expectation operation.
$\operatorname{Re}[\cdot]$ and $\operatorname{Im}[\cdot]$ represent the real and imaginary components of a vector or matrix, respectively.
For indexing, $\textbf{a}[k]$ and $\textbf{A}[i,j]$ denote the $k$-th element of a vector and the $(i,j)$-th entry of a matrix, respectively.
$\textbf{0}_{N\times M}$ and $\textbf{1}_{N\times M}$ represent the all-zero and all-one matrices of size $N \times M$, and $\textbf{I}_{N}$ represents the $N \times N$ identity matrix.

\section{System Model}\label{section2}
Consider a \gls{mimo} system where the transmitter employs a reconfigurable \gls{mamp} configuration consisting of $N_a$ active antennas (i.e., the number of ESPAR) and $N_p$ parasitic antennas, resulting in a total of $N = N_a + N_p$ transmit antennas, while the receiver uses an array consisting of $M$ active antennas. In prior studies on parasitic antenna arrays, beamforming was typically achieved by attaching varactors to the parasitic element and continuously tuning their reactances within a certain range \cite{barousis_beamspace-domain_2011, deshpande_beamforming_2025, zhang_analog_2023, bucheli_garcia_low-complexity_2020}. 
However, varactor-based tuning complicates the design of RF control system architecture in the RF circuitry, reducing both cost efficiency and operational stability.

\begin{figure}[t]
    \centering
    \includegraphics[width=0.43\textwidth]{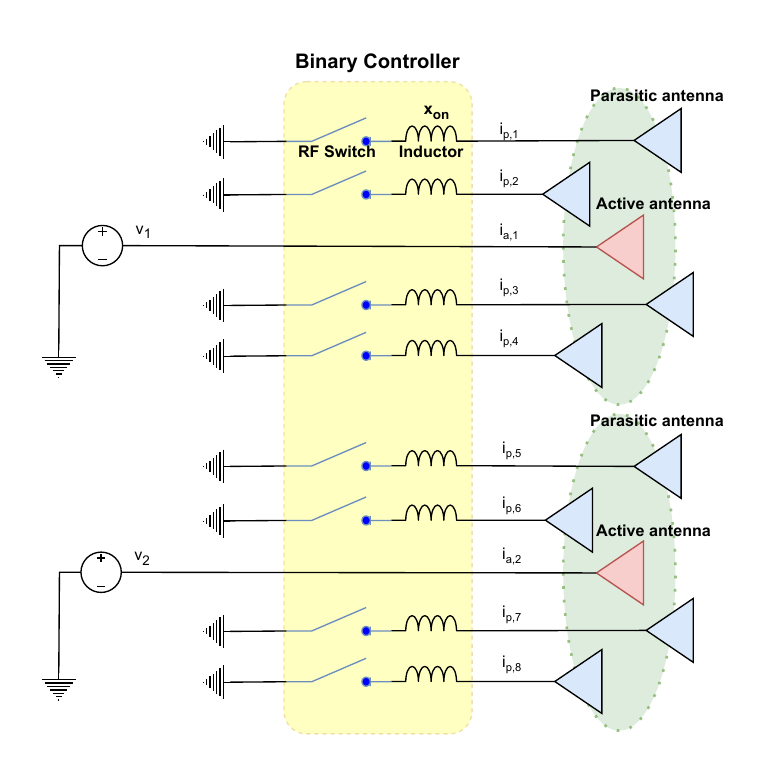}
    \caption{Proposed MAMP array with $N_a=2$ and $N_p=8$ }
    \label{fig:MAMP}
    \vspace{-5mm}
\end{figure}

In this work, we replace varactors with RF switches, such as PIN diodes or MEMS switches, to control the parasitic elements and propose a \gls{mamp} antenna system in which the parasitic elements are controlled through a simple binary controller composed of RF switches and lumped passive components, as shown in Fig.~\ref{fig:MAMP}. An RF switch exhibits a very small capacitance, corresponding to a large negative reactance, in the OFF state.
In the on state, an RF switch is designed to behave as a short circuit, although in practice it exhibits a finite reactance determined by the device packaging. Consequently, each parasitic antenna possesses only two possible reactance states, which are determined by the switch specifications, the operating frequency, and the fixed reactance element used in the ON state.

The primary objective of the proposed system is to develop a codebook-based transmit beamforming scheme for transmitters with a \gls{mamp} array. The design particularly addresses scenarios where channel reciprocity does not hold, such as in frequency-division duplexing (FDD) systems. 

\subsection{Beamspace Channel Model}
In modeling the channel of a parasitic antenna array, it is essential to consider the radiation pattern, since the channel characteristics are influenced by the on-off states of the parasitic elements and the geometry of antenna array. In the far-field, the radiation pattern can be expressed by the superposition of radiation patterns as \cite{sun_fast_2004}
\begin{equation}
\label{eq:radiation pattern}
P(\theta) = \sum_{n=1}^{N} i_n a_n(\theta) = \textbf{a}^T(\theta)\textbf{i}
\end{equation}
where \(\theta\) represents the azimuth angle in 2D coordinate system, and \( \textbf{a}(\theta) = [a_1(\theta),a_2(\theta),\dots,a_{N}(\theta)]^T \) is the steering vector at an angle \( \theta \).
Note that the steering vector is determined by the parasitic antenna array structure.

Accordingly, the channel can be represented using an angle-based virtual representation, as introduced in \cite{sayeed_deconstructing_2002}. Let $\textbf{H}_v \in \mathbb{C}^{K \times K}$ denote the virtual channel matrix, where each entry corresponds to the channel gain between a pair of uniformly quantized angle-of-arrival (AoA) and angle-of-departure (AoD) directions over $K$ angular bins.  
The physical channel matrix $\textbf{H} \in \mathbb{C}^{M \times N}$ can be expressed as
\begin{equation}
\label{eq:channel}
\textbf{H} = \sum_{i=1}^{K} \sum_{j=1}^{K} \textbf{H}_v[i,j] \, \textbf{a}_r(\theta_{r,i}) \, \textbf{a}_t^H(\theta_{t,j}) = \textbf{A}_r^H \textbf{H}_v \textbf{A}_t,
\end{equation}
where $\theta_{r,i}$ and $\theta_{t,j}$ denote the $i$-th AoA and $j$-th AoD, respectively.  
Here, $\textbf{a}_r(\theta_{r,i}) \in \mathbb{C}^{M \times 1}$ and $\textbf{a}_t(\theta_{t,j}) \in \mathbb{C}^{N \times 1}$ are the receiver and transmitter steering vectors, respectively.  
The matrices $\textbf{A}_r = [\textbf{a}_r(\theta_{r,1}), \dots, \textbf{a}_r(\theta_{r,K})]^T \in \mathbb{C}^{K \times M}$ and $\textbf{A}_t = [\textbf{a}_t(\theta_{t,1}), \dots, \textbf{a}_t(\theta_{t,K})]^T \in \mathbb{C}^{K \times N}$ are the receiver and transmitter steering matrices, respectively.  
Each steering matrix is constructed by stacking the corresponding steering vectors associated with uniformly quantized angular directions. The steering matrices project the physical domain signals into the virtual angular domain.

Using \eqref{eq:radiation pattern}, the radiation pattern vector \(\textbf{e}_t \in \mathbb{C}^{K \times 1}\) sampled over \(K\) angular bins at the transmitter, is given by 
\begin{equation}\label{eq:sampled radiation}
\textbf{e}_t(t) = \textbf{A}_t \textbf{i}(t) s(t),
\end{equation}
where \(s(t) \in \mathbb{C}\) denotes a continuous signal corresponding to the transmitted symbol $s$.
Accordingly, the received signal $\textbf{y}(t) \in \mathbb{C}^{M\times1}$ at time \(t\) can be written as
\begin{equation}\label{eq:received_signal_model}
\textbf{y}(t) = \textbf{A}_r^H \textbf{H}_v(t)\textbf{A}_t \textbf{i}(t) s(t) + \textbf{n}(t) = \textbf{H}(t)\textbf{i}(t)s(t) + \textbf{n}(t),
\end{equation}
where $\textbf{n}(t) \in \mathbb{C}^{M\times1}$ is assumed to be an i.i.d. circularly symmetric complex Gaussian random vector, i.e., $\textbf{n}(t) \sim \mathcal{CN}(\textbf{0}_{M \times 1}, \sigma_n^2 \textbf{I}_{M})$. Without loss of generality, we assume a narrowband block channel model where the channel remains constant for \(L\) symbol times \cite{love2008overview}.  
Throughout the remainder of this paper, the time index $t$ is omitted for notational simplicity.
As observed in \eqref{eq:received_signal_model}, the current vector $\textbf{i}$ functions similarly to a beamforming vector in the context of the parasitic antenna model.  
Hence, the subsequent sections are devoted to the design of the current vector.

\subsection{Representation of Current in MAMP}
The active antenna is equipped with a voltage source \( \textbf{v}_a = [v_1, v_2, \dots, v_{N_a}]^T \in \mathbb{C}^{N_a\times1} \), while the passive antennas are characterized by variable reactances \(\textbf{x} = [x_1, x_2, \dots, x_{N_p}]^T \in \mathbb{R}^{N_p\times1}\). Since the antenna spacing of the parasitic antenna array is smaller than \( \lambda/2 \), the mutual coupling allows the currents in each antenna to be induced by the voltage sources and variable reactances. Therefore, the current vector \( \textbf{i} =[\textbf{i}_a^T, \textbf{i}_p^T]^T \) can be written as
\begin{equation}
\label{eq:ESPAR_Current}
\textbf{i} = (\textbf{Z} + j\textbf{X})^{-1} \textbf{v},
\end{equation}
where \(\textbf{i}_a = [i_{a,1}, i_{a,2}, \dots, i_{a,N_a}]^T \in \mathbb{C}^{N_a\times1}\) is the current vector flowing in active antennas and \( \textbf{i}_p = [i_{p,1}, i_{p,2},\dots, i_{N_p}]^T \in \mathbb{C}^{N_p\times1} \) is the current vector flowing through passive antennas, \( \textbf{Z}\in\mathbb{C}^{N\times N} \) denotes the impedance matrix and \( {Z}_{i,j} \) denotes mutual coupling between the \(i\)th and \(j\)th antennas, \( \textbf{X} = \text{diag}([\textbf{0}_{N_a}^T, \textbf{x}^T]) \in \mathbb{R}^{N\times N}\) represents the variable reactance matrix, and \( \textbf{v} = [\textbf{v}_a^T,\textbf{0}_{N_p \times 1}^T] \in \mathbb{C}^{N\times1} \) is the voltage vector. 

In terms of voltage, \eqref{eq:ESPAR_Current} can be rewritten as \textbf{v} = (\textbf{Z} + j\textbf{X})\textbf{i}.
Considering that the mutual coupling matrix is symmetric, the mutual coupling matrix, \(\textbf{Z} \), can be expressed as \cite{kalis_parasitic_2014}
\begin{equation}
\label{eq:Mutual Coupling matrix of circular array}
\textbf{Z} =
\begin{bmatrix}
Z_{11} & Z_{12} & \cdots & Z_{1M} \\
Z_{21} & Z_{22} & \cdots & Z_{2M} \\
\vdots & \vdots & \ddots & \vdots \\
Z_{M1} & Z_{N2} & \cdots & Z_{MM}
\end{bmatrix}
=
\begin{bmatrix}
\textbf{Z}_{a}                 & \textbf{Z}_{ap}^T \\
\textbf{Z}_{ap} & \textbf{Z}_p
\end{bmatrix},
\end{equation}
where \( \textbf{Z}_a \in \mathbb{C}^{N_a \times N_a} \), \( \textbf{Z}_{ap} \in \mathbb{C}^{N_p \times N_a} \), and \( \textbf{Z}_p \in \mathbb{C}^{N_p \times N_p} \) represent the mutual coupling matrices between active antennas, active antennas and parasitic antennas, parasitic antennas, respectively. Defining \( \textbf{Z} + j\textbf{X} \) as \( \bar{\textbf{Z}} \), \( \bar{\textbf{Z}} \) is given by
\begin{equation}
\label{eq:Z+jX}
\bar{\textbf{Z}} =
\begin{bmatrix}
\textbf{Z}_{a}                 & \textbf{Z}_{ap}^T \\
\textbf{Z}_{ap}      & \textbf{Z}_p + j\operatorname{diag}(\textbf{x})
\end{bmatrix}.
\end{equation}
Using \eqref{eq:Mutual Coupling matrix of circular array}, \( \textbf{v} \) can be rewritten as
\begin{equation}
\label{eq:division of ESPAR equation}
\textbf{v}_a = \begin{bmatrix} \textbf{Z} &\textbf{Z}_{ap}^T\end{bmatrix}\textbf{i},  
\end{equation}
\begin{equation}\label{eq:division of ESPAR equation2}
    \textbf{0}_{N_p \times 1}= \begin{bmatrix} \textbf{Z}_{ap}      & \textbf{Z}_p + j\operatorname{diag}(\textbf{x})\end{bmatrix}\textbf{i}.
\end{equation}
From \eqref{eq:division of ESPAR equation2}, it can be observed that the current vector lies in the null space of \(\begin{bmatrix} \textbf{Z}_{ap}      & \textbf{Z}_p + j\operatorname{diag}(\textbf{x})\end{bmatrix}\).
Therefore, a valid current vector can be constructed as a linear combination of the basis vectors spanning this null space.
To realize this current at the antenna, the required voltage source can be computed using the matrix \(\begin{bmatrix} \textbf{Z} &\textbf{Z}_{ap}^T\end{bmatrix}\).
Using the matrix representation in \eqref{eq:Z+jX} to determine the basis, \( \textbf{i} \) can be expressed as
\begin{equation}
\label{eq:current from null space}
\textbf{i} =
\begin{bmatrix}
\textbf{i}_a \\
\textbf{i}_p
\end{bmatrix}
=\textbf{U}(\textbf{b})\textbf{i}_a,
\end{equation}
where
\begin{equation}
\label{eq:inverse_version_parasitic_current}
\textbf{U}(\textbf{b}) = 
\begin{bmatrix}
\textbf{I}_{N_a} \\
-(\textbf{Z}_p + j\operatorname{diag}(\textbf{x}))^{-1}\textbf{Z}_{ap}
\end{bmatrix},
\end{equation} 
and \( \textbf{i}_p = -(\textbf{Z}_p + j\operatorname{diag}(\textbf{x}))^{-1}\textbf{Z}_{ap}\textbf{i}_a\). From \eqref{eq:inverse_version_parasitic_current}, we can see that the induced current vector that flows through parasitic antennas is determined by \(\textbf{i}_a\) and \(\textbf{x}\).

The key distinction between conventional multiple antenna systems and \gls{mamp} lies in the presence of an inverse operation in the current computation.  
In conventional \gls{mimo}, mutual coupling is typically ignored, resulting in a linear relationship between the input voltage and the antenna currents \cite{li2017mimo}.  
In contrast, in \gls{mamp} systems, the antenna currents are calculated from both the input voltage and the variable reactances, where the reactances are embedded within a matrix inverse, as shown in \eqref{eq:ESPAR_Current}. This structure introduces significant computational complexity and complicates the design of beamforming strategies.
Therefore, in the following section, we aim to express the induced currents in the switched \gls{mamp} model without involving inverse terms.

\section{Parasitic Antenna Current Modeling and Parameter Extraction under ON-OFF Switching}\label{section3}
Many prior works on parasitic antenna systems have formulated the reactance vector $\textbf{x}$ as a continuous variable, typically implemented using varactors in RF circuitry. However, continuous tunable reactance introduces significant challenges in the design of control systems, requiring more complex and costly components and increasing the complexity of the control architecture. These challenges become more severe as the number of parasitic elements increases.
To address this issue, this paper proposes a parasitic antenna architecture with binary controllers, which can be easily realized through RF switches such as PIN diodes or MEMS switches. This binary approach simplifies the control architecture and improves cost efficiency, making it suitable for practical implementation.

\subsection{Mathematical Approximation of Induced Currents}
\label{subsection:approx_current}
In parasitic antenna arrays, the expression of the current vector $\textbf{i}$ requires a matrix inversion determined by the reactance vector $\textbf{x}$, as shown in \eqref{eq:ESPAR_Current}, which increases the computational complexity and complicates the overall problem. Consequently, prior works employing continuous reactance control attempted to linearize \eqref{eq:ESPAR_Current} using a first-order Taylor series approximation \cite{bucheli_garcia_low-complexity_2020,han_spectrally_2023}. In this paper, we mathematically model the induced currents in the binary-state parasitic antenna array in order to closely match the actual induced currents.

To represent the on-off state of parasitic antenna current, \(\textbf{i}_p\), we define two sets of selection matrices proposed in \cite{liu_sensor_2016}
\begin{equation}
\mathcal{S}_{\text{on}} =\bigl\{\bm{\Phi}\in\{0,1\}^{r\times (N_p)}\big|
\bm{\Phi}\bm{\Phi}^T = \textbf{I}_{r}, \bm{\Phi}^T\bm{\Phi} = \operatorname{diag}(\textbf{b})\bigr\},
\end{equation}
\begin{equation}
\begin{split}
\mathcal{S}_{\text{off}} 
= \bigl\{\bm{\Psi}\in\{0,1\}^{(N_p-r)\times (N_p)} \,\big|\,
& \;\bm{\Psi}\bm{\Psi}^T = \textbf{I}_{N_p-r}, \\
& \;\bm{\Psi}^T\bm{\Psi}=\textbf{I}_{N_p}-\operatorname{diag}(\textbf{b})
\bigr\},
\end{split}
\end{equation}
where \(r\) denotes the number of parasitic antennas that are turned ON and \( \textbf{b} \in \{0,1\}^{N_p} \) is a binary vector in which each entry is set to 1 if the corresponding parasitic antenna is ON, and 0 if it is OFF. Each row of \(\bm{\Phi}\) and \(\bm{\Psi}\) has exactly one entry equal to 1 and all others 0, so that \(\bm{\Phi}\) selects \(r\) distinct indices and \(\bm{\Psi}\) selects \(N_p - r\) distinct indices from \(N_p\).

Let $\textbf{i}_{\text{on}} = \bm{\Phi} \textbf{i}_p$ and $\textbf{i}_{\text{off}} = \bm{\Psi} \textbf{i}_p$ denote the currents flowing through parasitic antennas in the ON and OFF states, respectively. It is worth noting that, in practical RF circuits, an RF switch in the OFF state still exhibits a finite reactance, which induces a nonzero current; 
hence, $\textbf{i}_{\text{off}} \neq \textbf{0}_{(N_p - r) \times 1}$. Then, the parasitic current vector \( \text{i}_p \) from \eqref{eq:current from null space} can be reordered as \( \textbf{i}_p = [\textbf{i}_{\text{on}}^T, \textbf{i}_{\text{off}}^T]^T \) and represented as
\begin{equation}
\label{eq:division of on-off state}
\textbf{i}_p =- 
\begin{bmatrix}
\textbf{Z}_{p,\text{on}} + jx_{\text{on}}\textbf{I}_r & \textbf{Z}_{p,\text{on-off}} \\
\textbf{Z}_{p,\text{on-off}}^T & \textbf{Z}_{p,\text{off}} + jx_{\text{off}}\textbf{I}_{N_p - r}
\end{bmatrix}^{-1}
\begin{bmatrix}
\textbf{Z}_{ap,\text{on}} \\
\textbf{Z}_{ap,\text{off}}
\end{bmatrix}\textbf{i}_a,    
\end{equation}
where \( \textbf{Z}_{p,\text{on}} =\bm{\Phi}\textbf{Z}_p\bm{\Phi}^T \) and \( \textbf{Z}_{p,\text{off}} = \bm{\Psi}\textbf{Z}_p\bm{\Psi}^T \) represent the mutual coupling among the ON-state and OFF-state parasitic antennas, respectively, while \( \textbf{Z}_{p,\text{on-off}}=\bm{\Phi}\textbf{Z}_p\bm{\Psi}^T \) denotes the mutual coupling between the ON-state and OFF-state parasitic antennas. Similarly, \( \textbf{Z}_{ap,\text{on}}=\bm{\Phi}\textbf{Z}_{ap}\) and \( \textbf{Z}_{ap,\text{off}} = \bm{\Psi}\textbf{Z}_{ap} \) represent the mutual coupling between the active antenna and the ON-state and OFF-state parasitic antennas, respectively. $x_{\text{on}}$ and $x_{\text{off}}$ denote the reactance values of each parasitic antenna in the ON and OFF states, respectively.

We denote
\begin{equation}
    \textbf{M} := 
    \begin{bmatrix}
        \textbf{B} & \textbf{C}\\
        \textbf{D} & \textbf{E}
    \end{bmatrix}
    =
    \begin{bmatrix}
\textbf{Z}_{p,\text{on}} + jx_{\text{on}}\textbf{I}_r & \textbf{Z}_{p,\text{on-off}} \\
\textbf{Z}_{p,\text{on-off}}^T & \textbf{Z}_{p,\text{off}} + jx_{\text{off}}\textbf{I}_{N_p - r}
\end{bmatrix}.
\end{equation}
Then, we can express the inverse of M using the Schur complement as \cite{zhang2006schur}
\begin{equation} \label{eq:schur complement}
    \textbf{M}^{-1} = 
    \begin{bmatrix}
    \textbf{S}^{-1} & -\textbf{S}^{-1}\textbf{C}\textbf{E}^{-1}\\
    -\textbf{E}^{-1}\textbf{D}\textbf{S}^{-1} & \textbf{E}^{-1}+\textbf{E}^{-1}\textbf{D}\textbf{S}^{-1}\textbf{C}\textbf{E}^{-1}
    \end{bmatrix},
\end{equation}
where \( \textbf{S} = \textbf{B} - \textbf{C}\textbf{E}^{-1}\textbf{D}\). 
Since RF switches exhibit very small capacitance in the OFF-state, the corresponding reactance \( x_{\text{off}} \) is typically much larger than the entries of \( \textbf{B} \), \( \textbf{C} \), and \( \textbf{D} \), allowing us to reasonably assume \(\left\| \textbf{B}^{-1} \textbf{C} \textbf{E}^{-1} \textbf{D} \right\| < 1.\)

Under this condition, we can apply the Taylor series expansion to obtain
\( \textbf{S}^{-1} = \textbf{B}^{-1} + \sum_{n=1}^{\infty} \left( \textbf{B}^{-1} \textbf{C} \textbf{E}^{-1} \textbf{D} \right)^n \textbf{B}^{-1}\). To express \( \textbf{i}_{\text{on}}\) and \(\textbf{i}_{\text{off}}\) in polynomial forms, we need to compute \( \textbf{B}^{-1} \) and \( \textbf{E}^{-1} \), which can be written using the matrix inversion lemma as 
\begin{equation}\label{eq:inver_of_B_1st}
\begin{aligned}
    \textbf{B}^{-1} &= (\textbf{Z}_{p,\text{on}} + jx_{\text{on}}\textbf{I}_{r})^{-1} \\
    &=  \bm{\Phi}\bm{\Phi}^T(\bm{\Phi}\tilde{\textbf{Z}}_p\bm{\Phi}^T + \beta\bm{\Phi}\bm{\Phi}^T)^{-1}\bm{\Phi}\bm{\Phi}^T \\
    &= \bm{\Phi}[\tilde{\textbf{Z}}_p^{-1} -\tilde{\textbf{Z}}_p^{-1}(\tilde{\textbf{Z}}_p^{-1}+\beta^{-1}\operatorname{diag}(\textbf{b}))^{-1}\tilde{\textbf{Z}}_p^{-1}]\bm{\Phi}^T\\
    &=\bm{\Phi}[\tilde{\textbf{Z}}_p^{-1} -\tilde{\textbf{Z}}_p^{-1}(\textbf{I}_{M-1}+\beta^{-1}\tilde{\textbf{Z}}_p\operatorname{diag}(\textbf{b}))^{-1}]\bm{\Phi}^T,
\end{aligned}
\end{equation}
where \(\beta = jx_{\text{on}} +\textbf{Z}_p[i,i], \quad\forall i =1,2,\dots N_p\). Assuming all parasitic antennas have the same self impedance and ON-state reactance \(x_{\text{on}}\), \(\tilde{\textbf{Z}}_{p}\) is obtained by removing the diagonal elements from \(\textbf{Z}_p\). $|\textbf{Z}_p[i,i]|$ takes a value in the range of several tens of ohms up to approximately $100 ~\Omega$ in case of the dipole antenna~\cite{zhang2005dual}, and it is typically larger than the magnitudes of the off-diagonal elements of \(\tilde{\textbf{Z}}_{p}\). In addition, $x_{\text{on}}$ can be configured to have a reactance on the order of several tens of ohms by employing a fixed inductor, assuming that it behaves as an ideal lumped element in the sub-10GHz range. Accordingly, the norm \( \left\| \beta^{-1} \tilde{\textbf{Z}}_p \operatorname{diag}(\textbf{b}) \right\| < 1 \), and thus a Taylor series expansion can be applied to the inverse term of \eqref{eq:inver_of_B_1st}. Then we have
\begin{equation} \label{eq:inverse of A}
\begin{split}
\textbf{B}^{-1} 
&= \bm{\Phi}\Bigl[\beta^{-1}\operatorname{diag}(\textbf{b})
 - \beta^{-2}\operatorname{diag}(\textbf{b})\tilde{\textbf{Z}}_{p}\operatorname{diag}(\textbf{b}) \\
&\quad - \tilde{\textbf{Z}}_{p}^{-1}\sum_{n=3}^{\infty}
  (-\beta^{-1}\tilde{\textbf{Z}}_{p}\operatorname{diag}(\textbf{b}))^{n}\Bigr]\bm{\Phi}^T \\
&= \bm{\Phi}\Bigl[\beta^{-1}\operatorname{diag}(\textbf{b})
 - \beta^{-2}\operatorname{diag}(\textbf{b})\tilde{\textbf{Z}}_{p}\operatorname{diag}(\textbf{b})\Bigr]\bm{\Phi}^T \\ 
&\quad + \mathcal{O}(\beta^{-3}),
\end{split}
\end{equation}
where $\mathcal{O}(\epsilon)$ denotes a matrix-valued term whose 2-norm is 
bounded by $K\epsilon$ for some constant $K > 0$ as $\epsilon \to 0$.
The \( \mathcal{O}(\beta^{-3}) \) term can be safely neglected assuming that $|\beta|$ is sufficiently large. Following the same approach the inverse \( \textbf{E}^{-1} \) can be written as
\begin{equation}
    \begin{split} \label{eq:inverse of B}
        \textbf{E}^{-1} &= (\textbf{Z}_{p,\text{off}} + jx_{\text{off}}\textbf{I}_{N_p - r})^{-1} \\
        &=  \bm{\Psi}\bm{\Psi}^T(\bm{\Psi}\textbf{Z}_p\bm{\Psi}^T + jx_{\text{off}}\bm{\Psi}\bm{\Psi}^T)^{-1}\bm{\Psi}\bm{\Psi}^T\\
        &= \bm{\Psi}[\gamma^{-1}\operatorname{diag}(\textbf{1}_{N_p \times 1}-\textbf{b}) \\
        &\quad -\tilde{\textbf{Z}}_{p}^{-1}\sum_{n=2}^{\infty}(-\gamma^{-1}\tilde{\textbf{Z}}_{p}\operatorname{diag}(\textbf{b}))^{n}]\bm{\Psi}^T\\
        &= \bm{\Psi}[\gamma^{-1}\textbf{I}_{N_p} -\gamma^{-1}\operatorname{diag}(\textbf{b)}]\bm{\Psi}^T + \mathcal{O}\bigl(\gamma^{-2}\bigr),
    \end{split}
\end{equation}
where \(\gamma = jx_{\text{off}} +\textbf{Z}_p[i,i], \quad \forall i =1,2,\dots N_p\), and the OFF-state reactance \( x_{\text{off}} \) is typically set to several hundreds to thousands of ohms, depending on the specifications of the PIN diode or MEMS switch and the operating frequency. Assuming that \( |\gamma|\) is sufficiently large, the second-order term \( \mathcal{O}(\gamma^{-2}) \) can be safely neglected. 

Let \( \textbf{i}_{p,1} \) be the vector of size $N_p \times 1$ obtained by scaling and extending \( \textbf{i}_{\text{on}} \). Then, substituting \eqref{eq:inverse of A} and \eqref{eq:inverse of B} into \eqref{eq:division of on-off state}, we have
\begin{equation} \label{eq:On_State_Current}
\begin{split}
    \textbf{i}_{p,1} 
    &= \bm{\Phi}^T \textbf{i}_{\text{on}} \\
    &= - \bm{\Phi}^T 
    \Big( 
        \textbf{S}^{-1} \textbf{Z}_{ap,\text{on}} 
        - \textbf{S}^{-1} \textbf{C} \textbf{E}^{-1} \textbf{Z}_{ap,\text{off}} 
    \Big)\textbf{i}_a \\
    &= -\bm{\Phi}^T 
    \Big[
        \textbf{B}^{-1} \textbf{Z}_{ap,\text{on}} 
        - \textbf{B}^{-1} \textbf{C} \textbf{E}^{-1} \textbf{Z}_{ap,\text{off}} 
    \Big]\textbf{i}_a \\
    &\quad + \mathcal{O}(\delta^{-3}) + \mathcal{O}(\gamma^{-2}) \\
    &= - 
    \Big[ 
        \bm{\Phi}^T \textbf{B}^{-1} \bm{\Phi} \textbf{Z}_{ap} 
        - \bm{\Phi}^T \textbf{B}^{-1} \bm{\Phi} \textbf{Z}_p 
        \bm{\Psi}^T \textbf{E}^{-1} \bm{\Psi} \textbf{Z}_{ap} 
    \Big]\textbf{i}_a \\
    &\quad + \mathcal{O}(\delta^{-3}) + \mathcal{O}(\gamma^{-2}) \\
    &= - 
    \Big[ 
        \big( \beta^{-1} \operatorname{diag}(\textbf{b}) 
        + \beta^{-2} \operatorname{diag}(\textbf{b}) \tilde{\textbf{Z}}_p \operatorname{diag}(\textbf{b}) \big) \\
        &\qquad \times \big( \textbf{I}_{N_p} 
        - \textbf{Z}_p \big( \gamma^{-1} \textbf{I}_{N_p} 
        - \gamma^{-1} \operatorname{diag}(\textbf{b}) \big) \big) \textbf{Z}_{ap} 
    \Big]\textbf{i}_a \\
    &\quad + \mathcal{O}(\delta^{-3}) + \mathcal{O}(\gamma^{-2}) \\
    &\approx - 
    \Big[ 
        \beta^{-1} 
        \big( \operatorname{diag}(\textbf{Z}_{ap}\textbf{i}_a) 
        - \gamma^{-1} \operatorname{diag}(\textbf{Z}_p \textbf{Z}_{ap} \textbf{i}_a) \big) \textbf{b} \\
        &\qquad 
        - \beta^{-1} \operatorname{diag}(\textbf{b}) 
        \big( \beta^{-1}\tilde{\textbf{Z}}_p - \gamma^{-1} \textbf{Z}_p \big) 
        \operatorname{diag}(\textbf{Z}_{ap}\textbf{i}_a) \textbf{b} 
    \Big],
\end{split}
\end{equation}

\noindent
where \(\delta = \min(\beta,\gamma)\) and \(\mathcal{O}(\delta^{-3}) \) is negligible for sufficiently large \(|\delta|\). Similarly, \(\textbf{i}_{p,2} \in \mathbb{R}^{N_p \times 1}\) can be obtained by scaling and extending \( \textbf{i}_{\text{off}}\) as
\begin{equation} \label{eq:Off_State_Current}
\begin{split}
    \textbf{i}_{p,2} 
    &= \bm{\Psi}^T \textbf{i}_{\text{off}} \\
    &= - \bm{\Psi}^T 
    \Big[
        -\textbf{E}^{-1} \textbf{D} \textbf{S}^{-1} \textbf{Z}_{ap,\text{on}} 
        + \textbf{E}^{-1} \textbf{Z}_{ap,\text{off}} \\
        &\qquad + \textbf{E}^{-1} \textbf{D} \textbf{S}^{-1} \textbf{C} \textbf{E}^{-1} \textbf{Z}_{ap,\text{off}}
    \Big]\textbf{i}_a \\
    &=  
    -\Big[
        -\bm{\Psi}^T \textbf{E}^{-1} \textbf{D} \textbf{S}^{-1} \textbf{Z}_{ap,\text{on}} 
        + \bm{\Psi}^T \textbf{E}^{-1} \textbf{Z}_{ap,\text{off}}
    \Big]\textbf{i}_a \\
    &\quad + \mathcal{O}(\delta^{-3}) + \mathcal{O}(\gamma^{-2}) \\
    &= - 
    \Big[
        -\bm{\Psi}^T \textbf{E}^{-1} \bm{\Psi} \textbf{Z}_p \bm{\Phi}^T \textbf{B}^{-1} \bm{\Phi} \textbf{Z}_{ap} 
        + \bm{\Psi}^T \textbf{E}^{-1} \bm{\Psi} \textbf{Z}_{ap}
    \Big]\textbf{i}_a \\
    &\quad + \mathcal{O}(\delta^{-3}) + \mathcal{O}(\gamma^{-2}) \\
    &\approx -
    \Big[
        \gamma^{-1} \textbf{Z}_{ap}\textbf{i}_a \\
        &\qquad - \gamma^{-1} \Big( 
            \beta^{-1} \textbf{Z}_p \operatorname{diag}(\textbf{Z}_{ap}\textbf{i}_a) 
            + \operatorname{diag}(\textbf{Z}_{ap}\textbf{i}_a) 
        \Big) \textbf{b} \\
        &\qquad + \gamma^{-1} \beta^{-1} 
        \operatorname{diag}(\textbf{b}) \textbf{Z}_p \operatorname{diag}(\textbf{Z}_{ap}\textbf{i}_a) \textbf{b}
    \Big].
\end{split}
\end{equation}

Assuming that $\delta^{-3}$ and $\gamma^{-2}$ are sufficiently small, we have
\begin{equation}\label{eq:ip_quadratic}
\begin{aligned}
    \textbf{i}_p(i_a, \textbf{b}) &= \textbf{i}_{p,1} + \textbf{i}_{p,2} = \textbf{T}(\textbf{b})\textbf{Z}_{ap}\textbf{i}_a \\ 
    &= \Big(-\frac{1}{\gamma}\textbf{I}_{N_p} + \operatorname{diag}(\textbf{b})\textbf{P} + \frac{1}{\beta \gamma}\textbf{Z}_p\operatorname{diag}(\textbf{b}) \\
    &\quad+ \operatorname{diag}(\textbf{b})\textbf{Q}\operatorname{diag}(\textbf{b})\Big)\textbf{Z}_{ap}\textbf{i}_a
\end{aligned}
\end{equation}
where \(\textbf{P} = \frac{\beta-\gamma}{\beta\gamma}\textbf{I}_{N_p}+\frac{1}{\beta\gamma}\textbf{Z}_p\), \(\textbf{Q} = -\frac{1}{\beta^2}\tilde{\textbf{Z}}_p - \frac{2}{\beta\gamma}\textbf{Z}_p\), and $\textbf{T}(\textbf{b})$ can be written in quadratic form with respect to $\textbf{b}$. Expanding \(\textbf{T}(\textbf{b})\) as a quadratic form in \( \textbf{b} \), we obtain
\begin{equation}
    \begin{aligned}
        \textbf{T}(\textbf{b}) = \textbf{T}_0 + \sum_{k=1}^{N_p}b_k\textbf{T}_k + \sum_{k=1}^{N_p}\sum_{l=1}^{N_p}b_kb_l\textbf{T}_{kl},
    \end{aligned}
\end{equation}
where \(\textbf{T}_0 = -\frac{1}{\gamma}\textbf{I}_{N_p}\), \(\textbf{T}_k = \textbf{V}_k\textbf{P}+\frac{1}{\beta\gamma}\textbf{Z}_p\textbf{V}_k\), \(\textbf{T}_{kl} = \textbf{V}_k\textbf{Q}\textbf{V}_l\), and $\textbf{V}_k \in \mathbb{R}^{N_p \times N_p}$ is a diagonal matrix with a unit entry at the $k$-th diagonal position and zeros elsewhere. The current vector $\textbf{i}$ can also be expressed as a quadratic form in $\textbf{b}$, yielding
\begin{equation}\label{eq:quadratic_of_ia}
    \textbf{i} = \tilde{\textbf{U}}(\textbf{b})\textbf{i}_a = 
    \begin{bmatrix}
        \textbf{I}_{N_a} \\ \textbf{T}(\textbf{b}) \textbf{Z}_{ap}
    \end{bmatrix} \textbf{i}_a,
\end{equation}
where $\tilde{\textbf{U}}(\textbf{b})$ is a matrix consisting of quadratic terms in $\textbf{b}$. While \(\textbf{U}(\textbf{b})\) in \eqref{eq:inverse_version_parasitic_current} is expressed as an inverse form with respect to the binary vector, through an approximation, \(\tilde{\textbf{U}}(\textbf{b})\) can be reformulated as a quadratic function of the binary vector, enabling the development of low-complexity algorithmic approaches.
\begin{figure}[t]
  \centering
  \begin{subfigure}[b]{0.48\columnwidth}
    \centering
    \includegraphics[width=\linewidth]{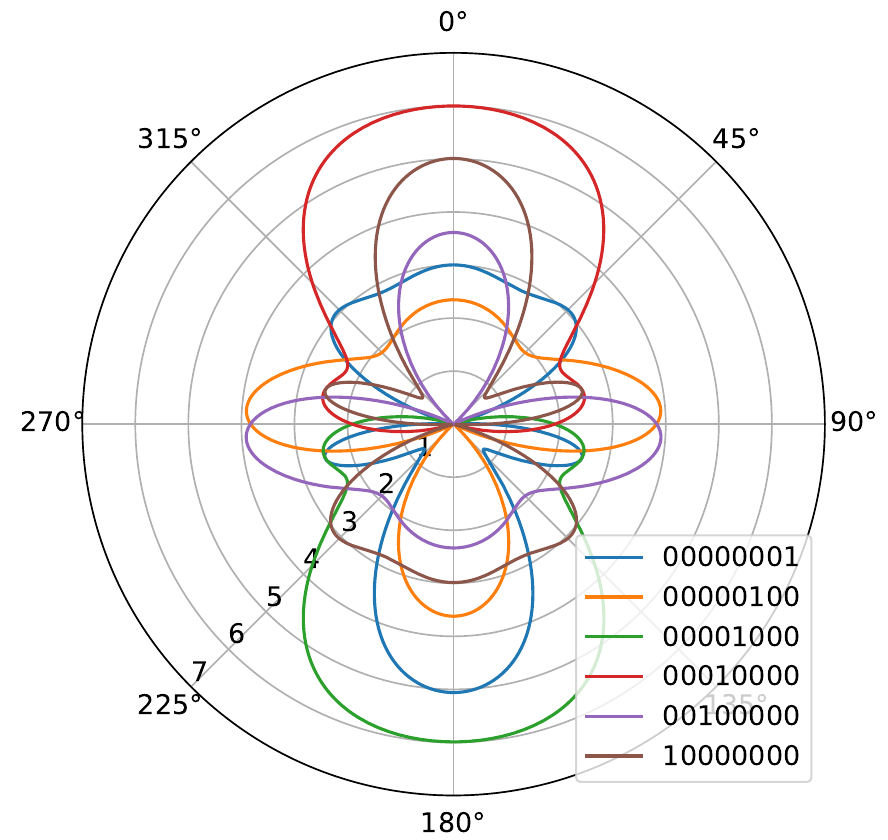}
    \caption{Linear array}
  \end{subfigure}
  \hfill
  \begin{subfigure}[b]{0.48\columnwidth}
    \centering
    \includegraphics[width=\linewidth]{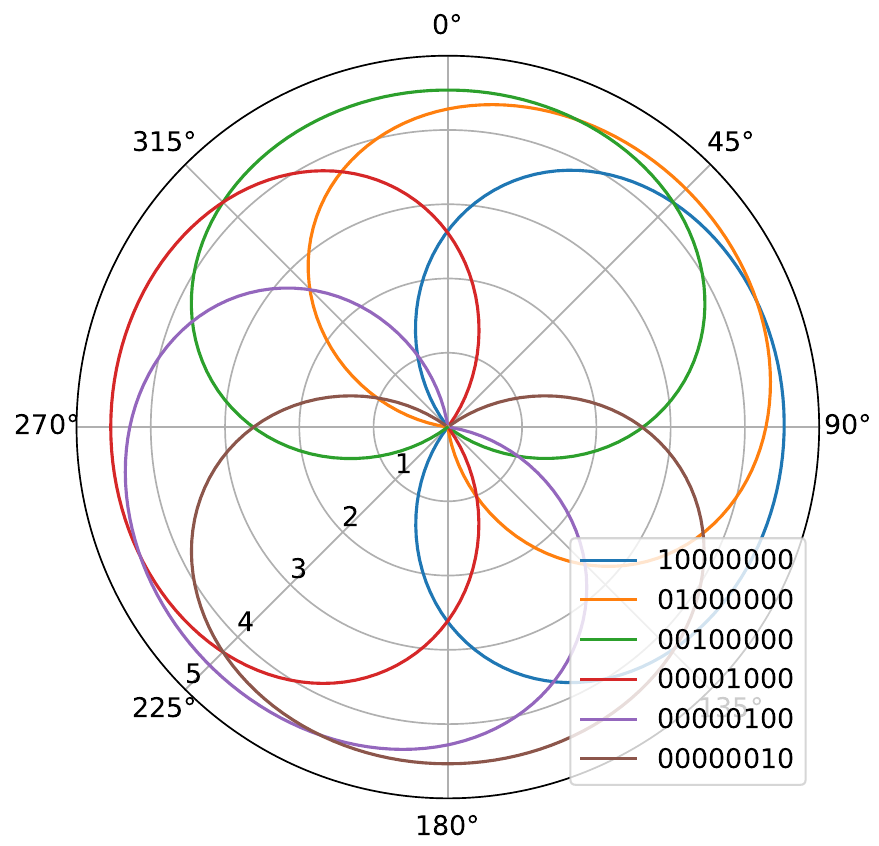}
    \caption{Circular array}
  \end{subfigure}
  \caption{Azimuth beam steering by binary control}
  \label{fig:array comparison}
  \vspace{-5mm}
\end{figure}

\begin{figure}[t]
    \centering
    \includegraphics[width=\linewidth]{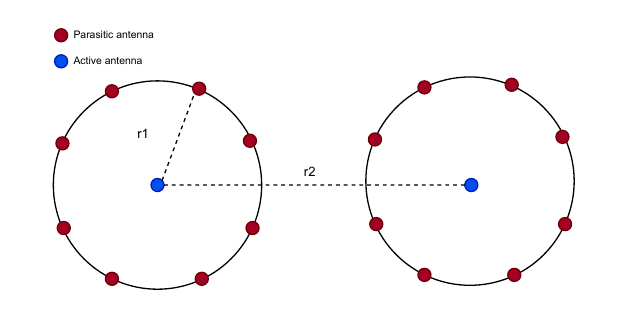}
    \caption{Configuration of MAMP Array}
    \label{fig:MAMP_Array}
    \vspace{-5mm}
\end{figure}

\begin{figure}[t]
    \centering
    \includegraphics[width=0.9\linewidth]{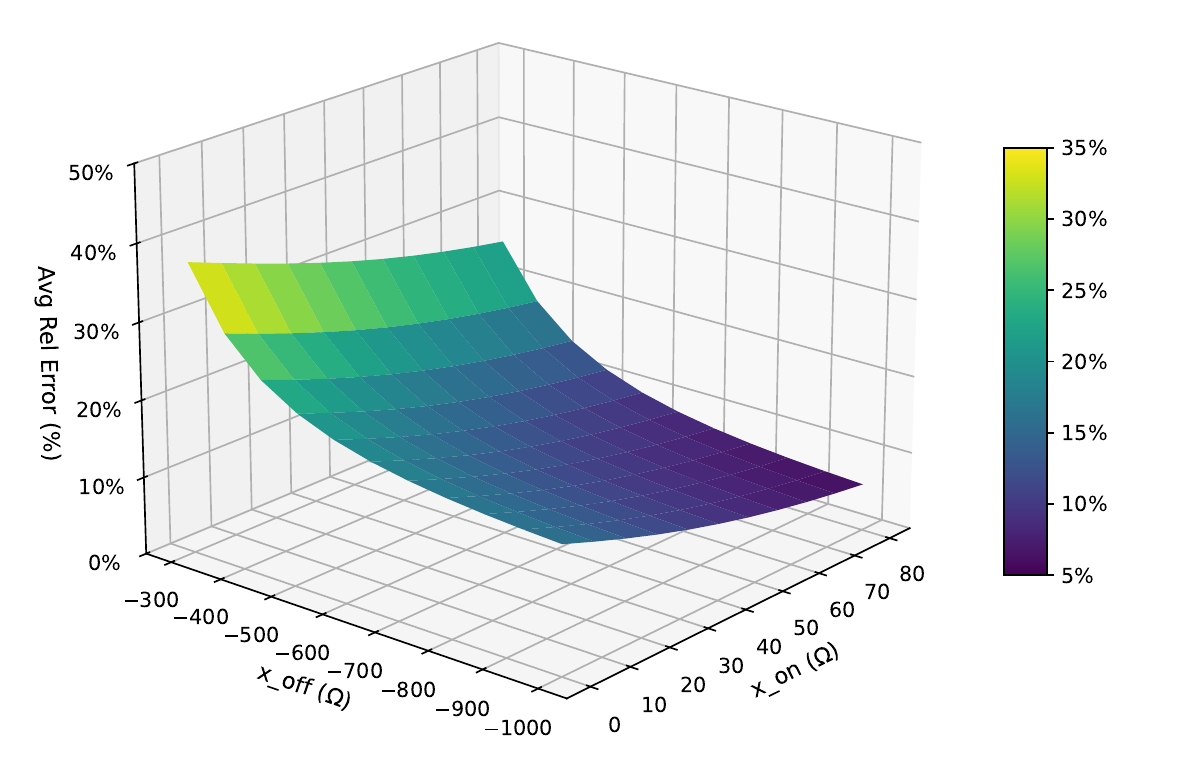}
    \caption{Average relative error versus $(x_{\mathrm{on}}, x_{\mathrm{off}})$}
    \label{fig:avg_relative_error}
    \vspace{-5mm}
\end{figure}
\begin{figure}[t]
    \centering
    \includegraphics[width=0.8\linewidth]{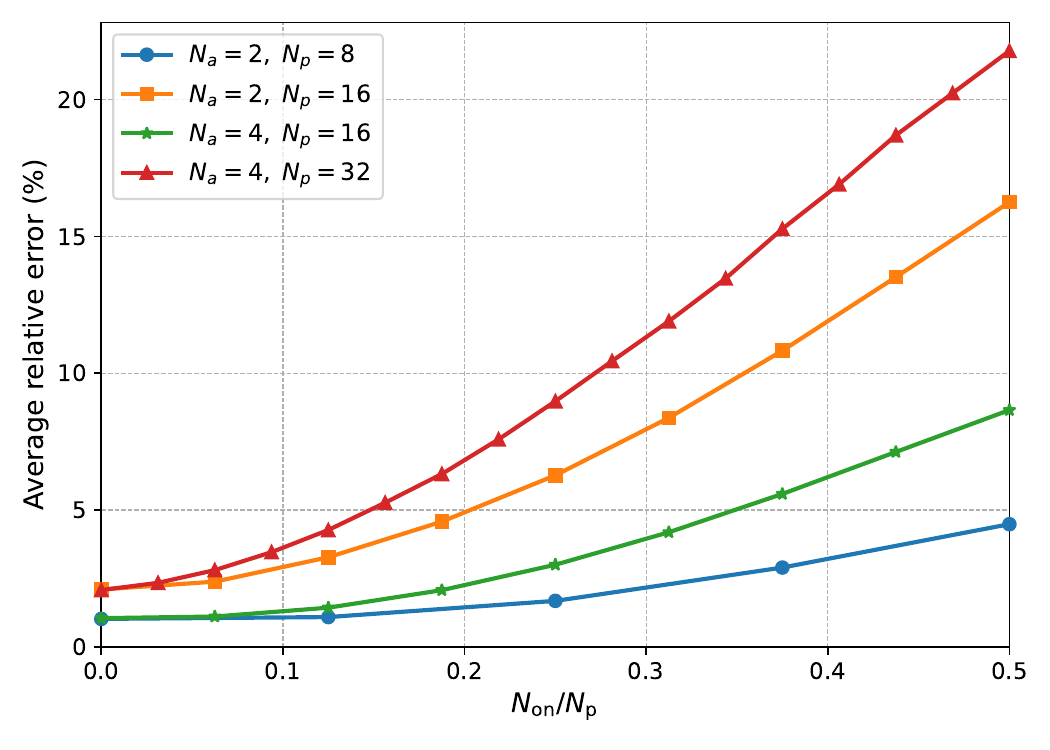}
    \caption{Average relative error across antenna configurations}
    \label{fig:avg_relative_error_diff_config}
    \vspace{-5mm}
\end{figure}

\subsection{Parameter Identification of \(x_{\text{on}}\) and \(x_{\text{off}}\)} 
\label{subsec:x_on and x_off}
In this section, we experimentally determine the array parameters required for the mathematical expression of the induced currents derived in the previous section to operate accurately. Each passive antenna can assume a value of \( x_{\text{on}} \) in ON-state and \( x_{\text{off}} \) in the OFF-state, depending on the configuration of the RF switch. In general, RF switches exhibit an OFF-state capacitance on the order of $60$--$200~\text{fF}$ \cite{hindle2010state}. Consequently, the corresponding reactance values vary depending on the switch specifications and the operating frequency.  

In contrast, when the RF switch is in the ON-state, it exhibits a small resistance, which may lead to energy loss. However, this effect is considered 
negligible and is therefore ignored in this work. The value of $x_{\text{on}}$ can be arbitrarily configured using a fixed inductor or capacitor, typically within the range of $0$ to $100~\Omega$. Note that large-magnitude reactance values reduce the power of the induced currents, resulting in significantly smaller power compared 
to the currents flowing through the active antennas, thereby limiting the diversity of the achievable beam patterns.

Let \( \textbf{i}_{\text{real}} \) as the actual current vector computed from \eqref{eq:ESPAR_Current}, and the relative error of the approximation is defined as
\begin{equation}\label{eq:relative_error}
    \text{error}(\tilde{\textbf{i}}) = \frac{ \left\| \textbf{i}_{\text{real}} - \tilde{\textbf{i}} \right\|_2 }{ \left\| \textbf{i}_{\text{real}} \right\|_2 }.
\end{equation}
This error is influenced by the terms \( \mathcal{O}(\gamma^{-2}) \) and \( \mathcal{O}(\delta^{-3}) \), 
which correspond to the effects of the diagonal term of \(\textbf{Z}_p \), \( x_{\text{on}} \) and \( x_{\text{off}} \). In addition, let the number of antennas in the ON-state be denoted by $N_{\text{on}}$. This parameter directly affects the dimension of the error matrix, and 2-norm of the error matrix increases as $N_{\text{on}}$ becomes larger.

The proposed signal processing framework is generally applicable to a wide range of antenna configurations. In this paper, we adopt a circular subarray structure for the proposed MAMP system due to its isotropic beam steering capability, rotational symmetry, and equal mutual coupling. A circular array enables beam steering over the entire azimuth plane through just controlling the parasitic antenna states while maintaining a direction-independent array response as shown in Fig.~\ref{fig:array comparison}. More importantly, all parasitic antennas are located at the same distance from the active antenna, yielding nearly identical mutual coupling between the active antenna and each parasitic element. When the parasitic states are represented by a binary control vector, this property makes the contribution of each binary element more uniform. Furthermore, the diagonal entries of the mutual coupling matrix become nearly identical, which improves the accuracy of the proposed quadratic approximation.

\begin{figure}[t]
  \centering
  \begin{subfigure}[b]{0.48\columnwidth}
    \centering
    \includegraphics[width=\linewidth]{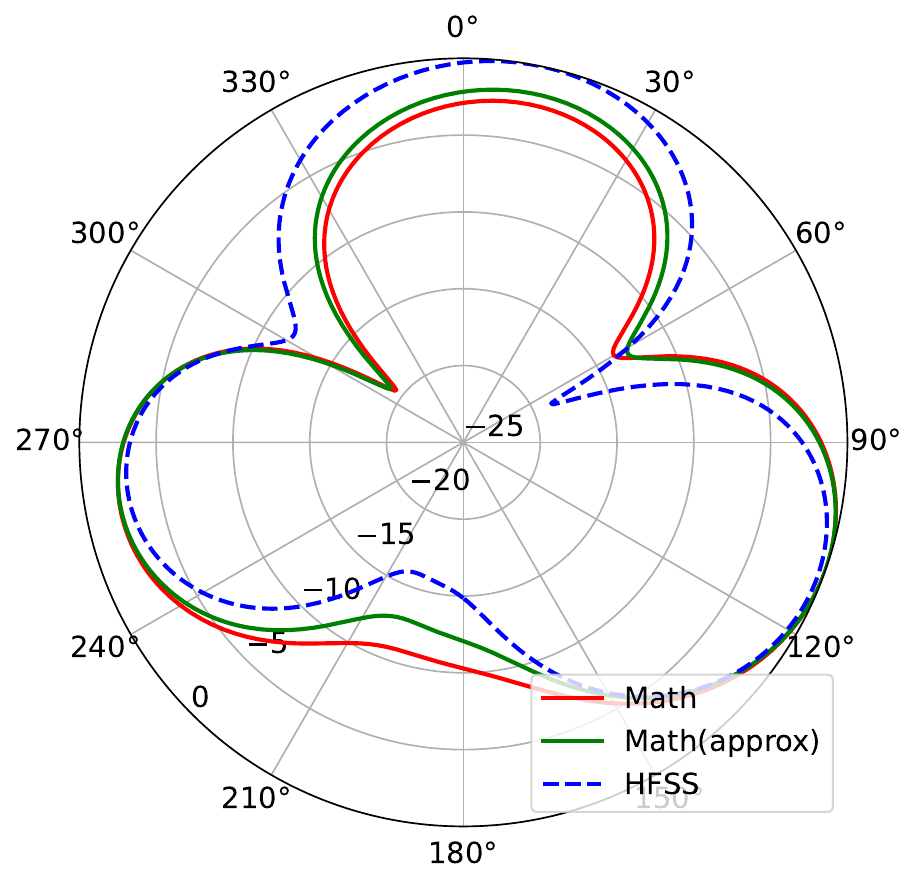}
    \caption{$r_1=0.25\lambda$, $r_2=0.6\lambda$}
    \label{fig:case1}
  \end{subfigure}
  \hfill
  \begin{subfigure}[b]{0.48\columnwidth}
    \centering
    \includegraphics[width=\linewidth]{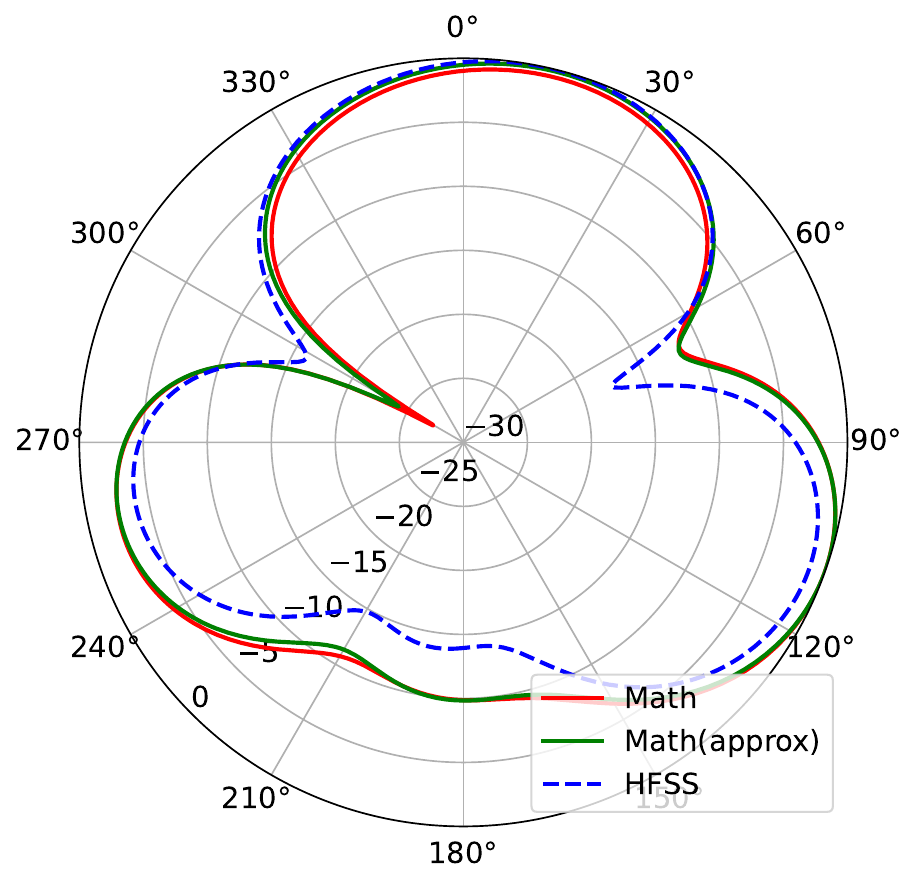}
    \caption{$r_1=0.25\lambda$, $r_2=0.7\lambda$}
    \label{fig:case2}
  \end{subfigure}
  \begin{subfigure}[b]{0.48\columnwidth}
    \centering
    \includegraphics[width=\linewidth]{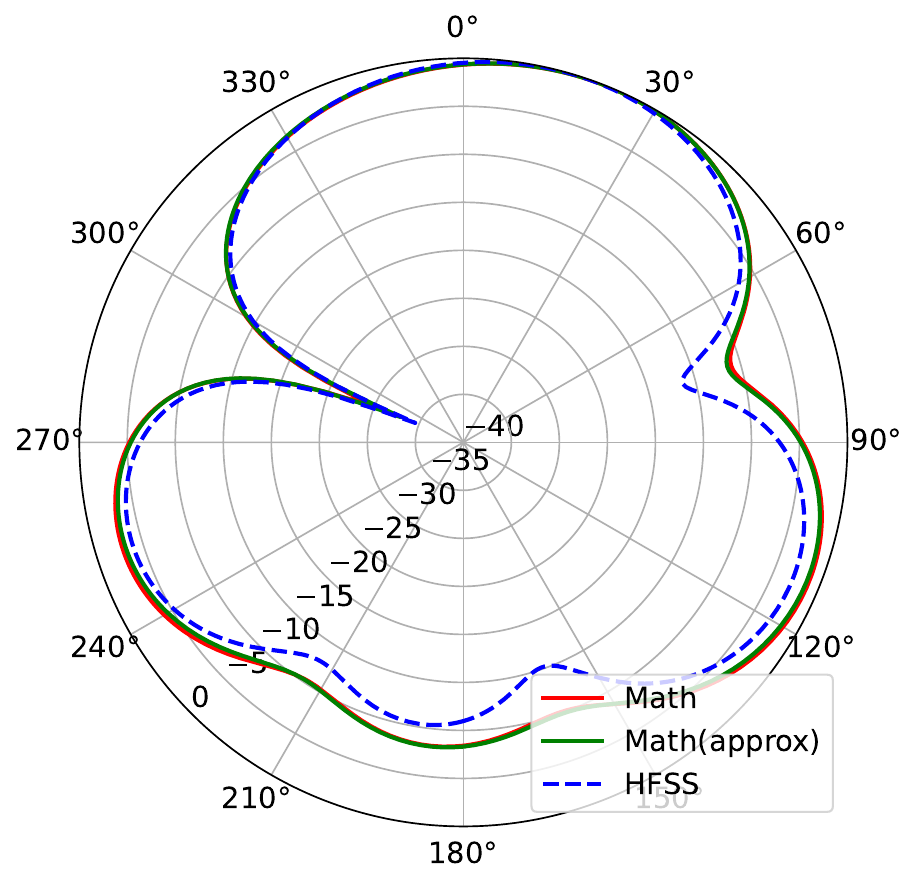}
    \caption{$r_1=0.3\lambda$, $r_2=0.8\lambda$}
    \label{fig:case3}
  \end{subfigure}
  \hfill
  \begin{subfigure}[b]{0.48\columnwidth}
    \centering
    \includegraphics[width=\linewidth]{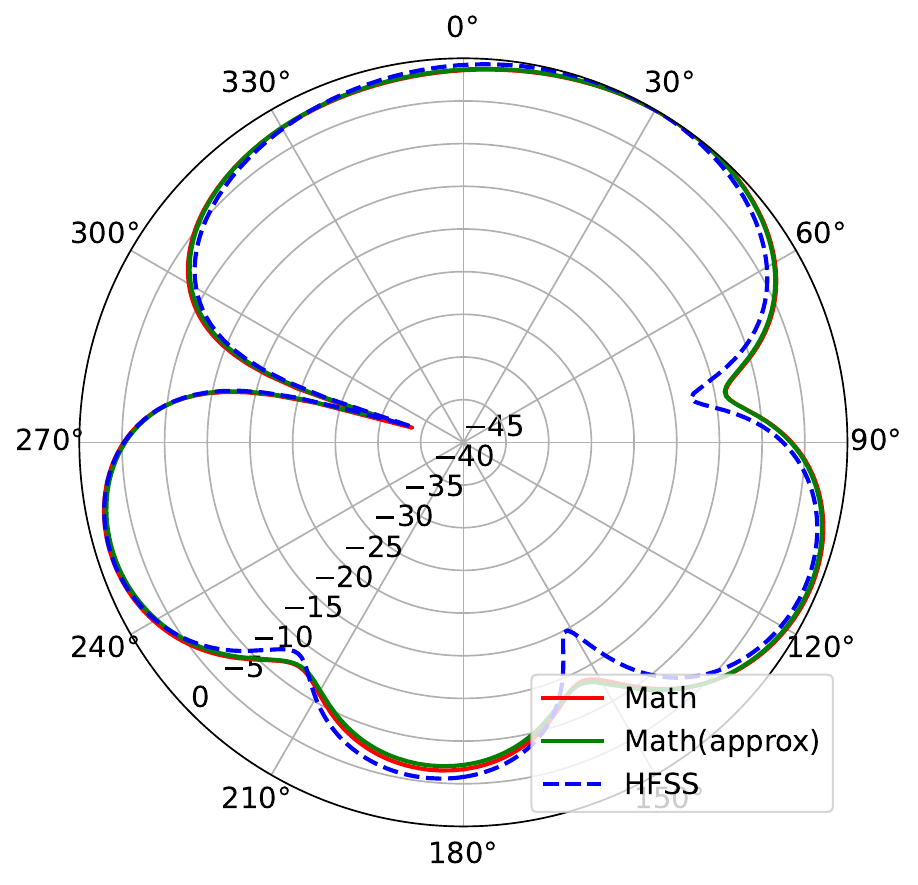}
    \caption{$r_1=0.3\lambda$, $r_2=0.9\lambda$}
    \label{fig:case4}
  \end{subfigure}

  \caption{Azimuth patterns for various ($r_1$, $r_2$) combinations with $N_{\text{on}}=1$, $\textbf{b}=[1,0,\dots,0]^T$, and relative phase = $90^\circ$.}
  \label{fig:multi_phase_patterns_N_on_1}
  \vspace{-5mm}
\end{figure}

The antenna array setup consists of \(N_a\) active antennas, each surrounded by equally spaced $N_p$ parasitic antennas forming a circular array shown in Fig.~\ref{fig:MAMP_Array}. The distance between adjacent active antennas and between each active antenna and its surrounding parasitic antennas are $0.9\lambda$ and $0.3\lambda$, respectively, where $\lambda$ denotes the 
wavelength. The rationale for adopting this antenna setup is to ensure that the far-field radiation patterns closely match both the mathematical expression in (1) and the \gls{hfss} simulation results, thereby enhancing the practicality of the study. The detailed experimental results are presented in the following section.

Using the relative error defined in (22) and the antenna setup, in Figure~\ref{fig:avg_relative_error}, we evaluated the accuracy of the approximate current in (21) by varying $x_{\text{off}}$ and $x_{\text{on}}$ when \(N_a =2, \ N_p =16\) and \(N_{\text{on}} \leq 4\). The range of $x_{\text{off}}$ was determined based on practical RF components. Since commercial RF switches exhibit an OFF-state capacitance of approximately $60$--$200~\text{fF}$ \cite{hindle2010state}, the corresponding reactance at $2.4~\text{GHz}$ was set to the range of $-1000$ to $-300~\Omega$. In the case of $x_{\text{on}}$, the diagonal terms of $\textbf{Z}_p$ obtained from \gls{hfss} simulations in the proposed setup exhibit positive reactance values. To reduce the error by increasing $\beta$ and $\gamma$, the range of $x_{\text{on}}$ was therefore set to $0$--$80~\Omega$, which can be readily implemented using an inductor. 
As shown in Fig.~\ref{fig:avg_relative_error}, the approximation error is approximately 5\% for \(x_{\text{on}} = 80\) and \(x_{\text{off}} = -1000\), which is sufficiently small to justify the proposed approximation. At \(2.4~\text{GHz}\), these reactance values correspond to an inductance of approximately 5.3~nH for \( x_{\text{on}} \) and a capacitance of approximately 66.3~fF for \( x_{\text{off}} \), which can be practically implemented using discrete inductors along with PIN diode switches or MEMS switches.

The impact of the $N_{\text{on}}$ constraint is illustrated in Fig.~5, which shows the average relative approximation error for various antenna configurations with different ratios of $N_{\text{on}}/N_p$
For most antenna configurations, the average relative approximation error remains well below $10\%$ when $N_{\text{on}}/N_p \leq 0.25$, suggesting that the approximation error is within an acceptable tolerance for the proposed beamforming design~\cite{bucheli_garcia_low-complexity_2020}. In addition, the $N_{\mathrm{on}}$ constraint significantly reduces the cardinality of the binary search space for large-scale codebook design while also simplifying the RF control circuit implementation. Therefore, it is adopted as a practical constraint in the proposed optimization problem.

\begin{figure}[t]
  \centering
  \begin{subfigure}[b]{0.48\columnwidth}
    \centering
    \includegraphics[width=\linewidth]{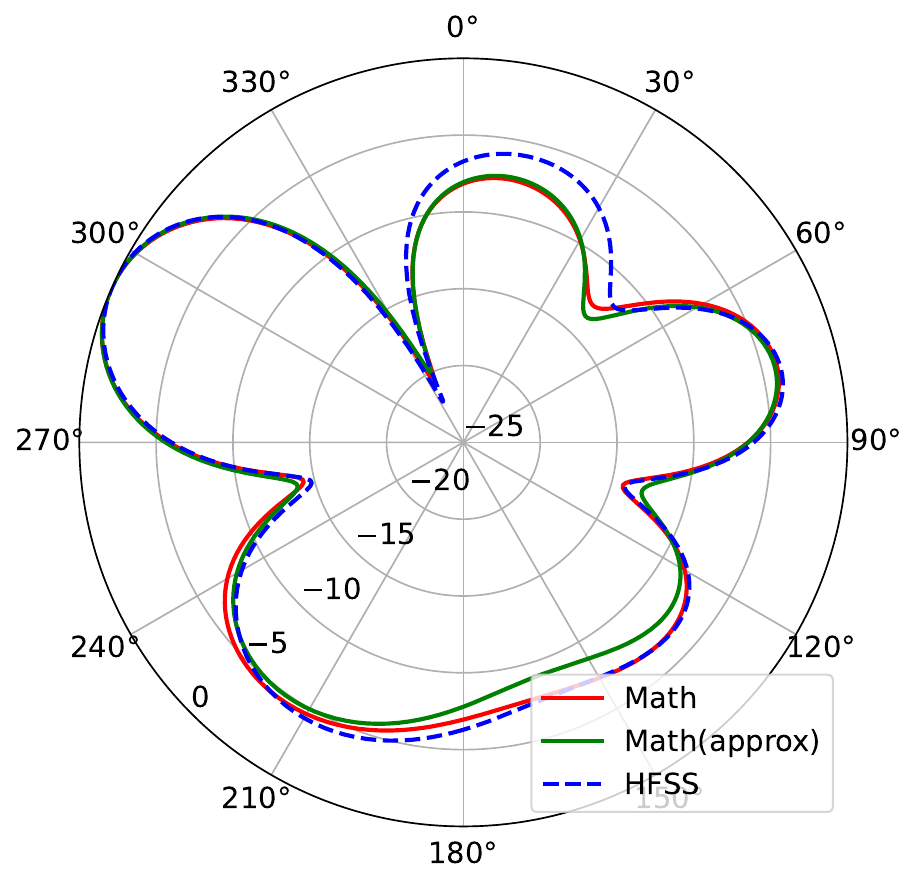}
    \caption{($N_a =2$, $N_p =16$, $N_{\text{on}}=4$)\\ $\textbf{b}[i] = 1$ for $i \in \{3, 6, 12,13\}$}
    \label{fig:case5}
  \end{subfigure}
  \hfill
  \begin{subfigure}[b]{0.48\columnwidth}
    \centering
    \includegraphics[width=\linewidth]{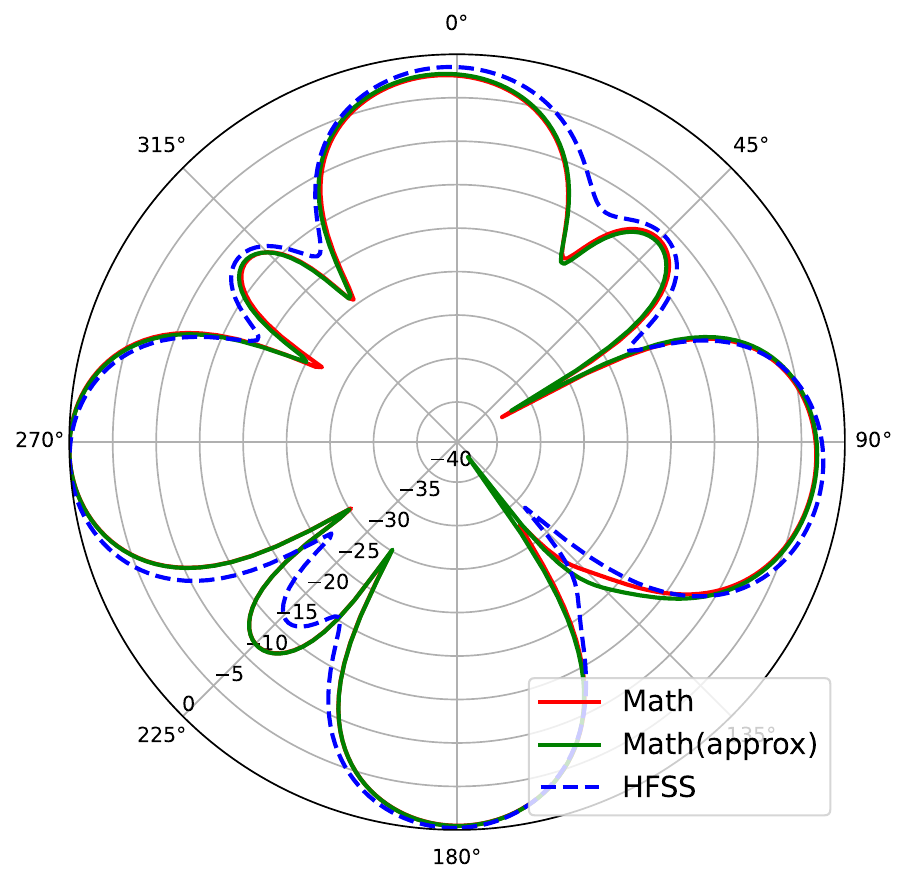}
    \caption{($N_a =4$, $N_p =16$, $N_{\text{on}}=4$)\\ $\textbf{b}[i] = 1$ for $i \in \{5,7,10,14\}$}
    \label{fig:case6}
  \end{subfigure}
  
  \caption{Azimuth radiation patterns for different antenna configurations with $r_1 = 0.3\lambda$ and $r_2 = 0.9\lambda$.}
  \label{fig:multi_phase_patterns}
  \vspace{-5mm}
\end{figure}

\subsection{HFSS-Based Validation of Radiation Patterns}\label{subsec: array parameters}
It is well known that the ideal array factor obtained solely from the element positions differs from the actual radiation pattern due to element patterns and mutual coupling \cite{balanis_antenna_2016}. Therefore, the radiation patterns derived from the proposed model should be validated against full-wave electromagnetic simulations, such as \gls{hfss}. 

In this section, we experimentally determine the array parameters by varying the radius of the circular arrays, denoted by $r_1$, and the spacing between the active antennas, denoted by $r_2$, and comparing the \gls{hfss} results with the mathematical radiation patterns obtained by substituting \eqref{eq:current from null space} and \eqref{eq:quadratic_of_ia} into \eqref{eq:sampled radiation}. For the simulations, two active antennas and sixteen parasitic antennas were employed. All antennas were dipole elements with a length of $0.5\lambda$, a wire radius of $0.3~\text{mm}$, and a gap spacing of $1~\text{mm}$ between the two arms of an individual dipole antenna. The same structure was used for both the active and parasitic elements. The RF switch is modeled as a two-state device, using the values obtained in \ref{subsec:x_on and x_off}. In the ON-state, it is represented by a $5.3~\text{nH}$ inductor corresponding to $x_{\text{on}} = 80~\Omega$ at $2.4~\text{GHz}$, while in the OFF-state, it is represented by a $66.3~\text{fF}$ capacitor corresponding to $x_{\text{off}} = -1000~\Omega$ at $2.4~\text{GHz}$.

Fig.~\ref{fig:multi_phase_patterns_N_on_1} and \ref{fig:multi_phase_patterns} present a comparison between full-wave simulation results obtained using \gls{hfss} and the analytical results, demonstrating the practical applicability of our approach in modeling realistic beam radiation patterns. The radiation patterns are plotted in the azimuth plane, with the maximum gain normalized to 0~dB. Each result corresponds to a radiation pattern obtained under a specific antenna array configuration, characterized by a binary state vector $\textbf{b}$ and a relative phase, defined as the phase difference between the excitation voltages $v_{0}$ and $v_{1}$, with the phase of $v_{0}$ fixed at $0^\circ$. 

Figs.~\ref{fig:multi_phase_patterns_N_on_1}(\subref{fig:case1}), 
(\subref{fig:case2}), (\subref{fig:case3}), and (\subref{fig:case4}) 
illustrate the radiation patterns when a single parasitic antenna is 
activated ($N_{\text{on}}=1$). In particular, the case shown corresponds to the binary vector $\textbf{b} =[1,0,\dots, 0]$ with a relative phase of $90^\circ$. This configuration serves as the baseline test case for analyzing the impact of $r_{1}$ and $r_{2}$ on the resulting radiation patterns.
In Fig.~\ref{fig:multi_phase_patterns_N_on_1}(\subref{fig:case1}), with $r_{1} = 0.25\lambda$ and $r_{2} = 0.6\lambda$, discrepancies are observed across most angular regions except near $120^\circ$ and $270^\circ$. 
The difference ranges from 2--3 dB up to 5 dB in certain regions. 
In Fig.~\ref{fig:multi_phase_patterns_N_on_1}(\subref{fig:case2}), 
the mathematical pattern consistently exceeds the HFSS pattern in the angular sector from $60^\circ$ to $270^\circ$, with discrepancies ranging from 1~dB to 4~dB.
In Fig.~\ref{fig:multi_phase_patterns_N_on_1}(\subref{fig:case3}), with $r_{1} = 0.3\lambda$ and $r_{2} = 0.8\lambda$, discrepancies are mainly observed in the range of $120^\circ$ to $220^\circ$ and around $75^\circ$, but these are  smaller compared to previous cases. However, in Fig.~\ref{fig:multi_phase_patterns_N_on_1}(\subref{fig:case4}), the discrepancy is almost negligible. This observation confirms that the discrepancies become markedly smaller as $r_{1}$ and $r_{2}$ increase. Through the baseline test, i.e., the simplest test case, we confirmed that the mathematical radiation pattern can be aligned with the practical pattern obtained from HFSS when $r_{1}=0.3\lambda$ and $r_{2}=0.9\lambda$. 

The validity of the selected values of $r_1$ and $r_2$ should also be verified for different configurations.
Fig.~7 compares the azimuth radiation patterns for $(N_a,N_p,N_{\mathrm{on}})=(2,16,4)$ and $(4,16,4)$. Although slight discrepancies can be observed in both cases, they are negligible. Furthermore, it can be observed that larger values of $N_{\text{on}}$ enable more dynamic beam patterns. The negligible difference between the red, green, and blue curves confirms that the formulas and parameter sets derived in \ref{subsection:approx_current} and \ref{subsec:x_on and x_off} provide practical values.

\section{Codebook Design Based on Exhaustive Searching}
\label{section4}
Based on the previously derived system model, we propose a practical transmit beamforming technique under a limited feedback architecture. 
As we mentioned earlier, the proposed MAMP model with on--off RF switching offers significant advantages in terms of stability and controllability compared to prior studies employing continuous reactance controllers. Compared to conventional \gls{mimo} systems composed solely of active antennas, the proposed framework can achieve the required performance with fewer active elements, thereby reducing the number of RF chains. Since RF chains are among the most power-hungry and costly components in antenna arrays, this makes the proposed scheme highly suitable for compact and energy-efficient transmitters.

Moreover, when combined with a codebook-based limited feedback system in environments where channel reciprocity is not guaranteed (e.g., FDD systems) \cite{love2008overview}, the transmitter is further simplified since no channel estimation procedure is required at the transmitter side. 
Furthermore, when a closed-form solution to the optimization problem is not available, the use of quantized inputs such as binary or other quantization schemes significantly reduces the search space compared to continuous values. As a result, a codebook can be readily constructed through exhaustive search.

\begin{algorithm}[!t]
\caption{Exhaustive Searching Codebook Design}\label{alg:proposed}
\begin{algorithmic}[1]
\STATE Construct \( \mathcal{I} = \{(\textbf{b}_1^{(0)}, \textbf{i}_{1}^{(0)}), \dots, (\textbf{b}_{2^B}^{(0)}, \textbf{i}_{2^B}^{(0)})\} \), which is the initial codebook consisting of binary and current vectors.
\REPEAT
    \STATE \( t = t +1\)
    \STATE (The Nearest Neighbor Rule) : Assign each channel vector \( \textbf{H} \) to regions \( \mathcal{A}^{(t)}_1, \mathcal{A}^{(t)}_2, ... , \mathcal{A}^{(t)}_{2^B} \) using \eqref{eq:nearest_neighbor_rule}
    \STATE (The Centroid Condition):
    \FORALL{ \(\textbf{b} \in \{\textbf{b}|\textbf{b}^T\textbf{b}\leq N_{\text{on}}\} \)}
        \STATE Solve GEVP and find \( \mu_{\max} \) using \eqref{eq:gen_eig}.
    \ENDFOR
    \STATE Select \( \textbf{b}_\ell^{(t)} \) and \(\textbf{u}_{l}^{(t)}\) to maximize \( \mu_{\max}\)
    \STATE Compute the active current \( \textbf{i}_{a,\ell}^{(t)} \) using \eqref{eq:ia_opt}
    \STATE Reconstruct the full current \( \textbf{i}_\ell^{(t)} \) using \eqref{eq:current from null space}
\UNTIL{ \( \max\limits_{\ell} \left\| \textbf{i}_\ell^{(t)} - \textbf{i}_\ell^{(t-1)} \right\| < \epsilon \) \quad \text{or} \quad \( t < N_{\text{iter}} \)}
\STATE Generate the final codebook \( \mathcal{B} = \{(\textbf{v}_{a,\ell}, \textbf{b}_\ell)\} \) from \( \mathcal{I} \) by computing each $\textbf{v}_a$ using \eqref{eq:division of ESPAR equation}
\end{algorithmic}
\end{algorithm}

To accurately evaluate the proposed modeling framework and beamforming algorithm while isolating the effects of other practical impairments, we assume that the receiver has access to both the channel distribution and perfect channel state information. Furthermore, the feedback link between the receiver and transmitter is assumed to be lossless and ideal. Although the induced current $\textbf{i}$ serves as the beamforming weight in our model, it is not suitable for direct use as a codebook. This is because the current is not directly controllable at the transmitter due to mutual coupling, and its relationship with the input voltage varies depending on the on-off states of the parasitic antennas.
Therefore, we propose a codebook \( \mathcal{B} = \{(\textbf{v}_a, \textbf{b})\}\) constructed from pairs of input voltage vectors and binary control vectors. While the problem is initially approached from the perspective of current-based beamforming weights, the 
ultimate goal is to identify the corresponding binary vectors and voltage sources that realize them. 

As mentioned earlier, the current vector serves as the transmit beamforming vector. In addition, one of the key considerations in transmit beamforming is the power constraint. In the proposed model, the transmit power from a circuit-theoretic perspective is defined  as \cite{ivrlac_toward_2010}
\begin{equation}\label{eq:PowerConstraints}
    P_{\text{tx}} = \textbf{i}^H \operatorname{Re}\{\textbf{Z}\} \textbf{i},
\end{equation}
and is later used as one of the constraints for optimizing the current vector.

With \( B \) feedback bits, we now quantize the channel vector \( \textbf{H} \) into \( 2^B \) regions, denoted as \( \mathcal{A}_1, \mathcal{A}_2, \ldots, \mathcal{A}_{2^B} \) and make the codebook, denoted as \(\mathcal{B} = [(\textbf{v}_1, \textbf{b}_1), ... , (\textbf{v}_{2^B}, \textbf{b}_{2^B})] \), aiming to design the quantizer and the codebook that maximize the received SNR for a given channel input. 
Codebook design for transmit beamforming under a unit power constraint, based on \gls{gla}, has been previously proposed in \cite{narula_efficient_1998, pengfei_xia_design_2006}. 
While this work also adopts the GLA framework, the characteristics of the parasitic antenna array require a modification of the power constraint, as described in \eqref{eq:PowerConstraints}. Accordingly, the proposed approach aims to maximize the SNR with respect to both \( \textbf{i}_a \) and \( \textbf{b} \), and the corresponding excitation voltage \( \textbf{v}_a \) can be readily computed.
The GLA framework requires the satisfaction of two key conditions: the centroid condition and the nearest neighbor condition.

Under the centroid condition, assuming perfect channel knowledge at the receiver and the use of maximal ratio combining (MRC), an optimization problem maximizing \gls{snr} can be formulated as \cite{lau2004design}
\begin{equation}
\begin{aligned}
    \textbf{i}_{l}^{\text{opt}} = \arg\max_{\textbf{i}} \quad & \textbf{i}^H \textbf{R}_\ell \textbf{i} \\
    \text{subject to} \quad & \textbf{i}^H \textbf{Z}_{\text{re}} \textbf{i} = P_{\max},
\end{aligned}
\label{eq:centroid_condition_opt}
\end{equation}
where \( \textbf{R}_\ell = \mathbb{E}[ \textbf{H}^H \textbf{H} \mid \textbf{H} \in \mathcal{A}_\ell ] \) is the conditional channel covariance matrix over region \( \mathcal{A}_\ell \), \( \textbf{Z}_{\text{re}} = \operatorname{Re}[\textbf{Z}] \) is real part of \(\textbf{Z}\) , and \( P_{\max} \) denotes the maximum allowable transmit power. 

By substituting the current vector with \eqref{eq:inverse_version_parasitic_current} and incorporating the binary vector constraint introduced in Section~\ref{subsec:x_on and x_off}, which accounts for approximation errors and limits the number of feasible combinations due to the computational complexity of codebook generation, the above formulation leads to the following joint optimization problem.
\begin{equation}
\begin{aligned}
    \textbf{i}^{\text{opt}}_{a,\ell}, \textbf{b}_{\ell} = \arg\max_{\textbf{i}_a, \textbf{b}} \quad & \textbf{i}^H_a \textbf{R}_{\ell,\text{eff}}(\textbf{b}) \textbf{i}_a \\
    \text{subject to} \quad &\textbf{i}_a^H \textbf{Z}_{\text{re},\text{eff}} \textbf{i}_a \leq P_{\max},
    \\& \textbf{b}^T\textbf{b} \leq N_{\text{on}},
\end{aligned}
\label{eq:centroid_condition_opt2}
\end{equation}
where \( \textbf{R}_{\ell,\text{eff}} = \textbf{U}(\textbf{b})^H\textbf{R}_{\ell}\textbf{U}(\textbf{b})\), \(\textbf{Z}_{\text{re},\text{eff}} = \textbf{U}(\textbf{b})^H\textbf{Z}_{\text{re}}\textbf{U}(\textbf{b})\), and \(N_{\text{on}}\) is a constraint on the number of ON-state antennas.

Since a joint optimization problem can be equivalently formulated as a nested optimization problem, especially when one variable block can be optimized given the other \cite{boyd_convex_2023}, the problem can be reformulated as a nested optimization
\begin{equation}\label{eq:centroid_condition_opt3}
\max_{\substack{\textbf{b} \in \{0,1\}^{N_p} \\ \textbf{b}^T \textbf{b} \leq N_{\text{on}}}} 
\left\{
    \max_{\substack{\textbf{i}_a \\ \textbf{i}_a^H \textbf{Z}_{\text{re},\text{eff}}(\textbf{b}) \textbf{i}_a \leq P_{\max}}} 
    \textbf{i}_a^H \textbf{R}_{\ell,\text{eff}}(\textbf{b}) \textbf{i}_a
\right\},
\end{equation}
where \( \textbf{i}_a \) is the variable of the inner optimization and \( \textbf{b} \) is the variable of the outer optimization. In such a nested formulation, if the inner optimization admits a closed-form solution with respect to the outer parameter, the problem can be solved by first determining the outer parameter and subsequently obtaining the inner parameter.

We first focus on the inner optimization in \eqref{eq:centroid_condition_opt3}. The optimal active current vector $\textbf{i}_a$ is obtained by solving
\begin{equation}
\max_{\textbf{i}_a}\ \textbf{i}_a^{H}\textbf{R}_{\ell,\mathrm{eff}}(\textbf{b})\textbf{i}_a
\quad \text{s.t.}\quad
\textbf{i}_a^{H}\textbf{Z}_{\mathrm{re,eff}}(\textbf{b})\textbf{i}_a \le P_{\max}.
\label{eq:inner_opt}
\end{equation}
By forming the Lagrangian and applying the \gls{kkt} conditions, the optimal solution satisfies the \gls{gevp}
\begin{equation}
\textbf{R}_{\ell,\mathrm{eff}}(\textbf{b})\,\textbf{i}_a
= \mu\, \textbf{Z}_{\mathrm{re,eff}}(\textbf{b})\,\textbf{i}_a.
\label{eq:gen_eig}
\end{equation}
Therefore, the optimal direction $\textbf{i}_a^{\text{opt}}$ is given by the generalized eigenvector corresponding to the maximum eigenvalue $\mu_{\max}$. The scaling is determined by the power constraint as
\begin{equation}\label{eq:ia_opt}
\textbf{i}_{a,\ell}^{\text{opt}}
= \sqrt{\frac{P_{\max}}
{\textbf{u}^H \textbf{Z}_{\mathrm{re,eff}}(\textbf{b})\textbf{u}}}
\textbf{u},
\end{equation}
where $\textbf{u}$ denotes the principal generalized eigenvector of $(\textbf{R}_{\ell,\mathrm{eff}}(\textbf{b}), \textbf{Z}_{\mathrm{re,eff}}(\textbf{b}))$. By substituting the result of the inner optimization into \eqref{eq:centroid_condition_opt3}, the outer optimization becomes
\begin{equation}\label{eq:centroid_condition_opt4}
\max_{\textbf{b} \in \{0,1\}^{N_p}} \quad \mu_{\text{max}}\Big(\textbf{R}_{l,\text{eff}}(\textbf{b})^{-1}\textbf{Z}_{l,\text{eff}}(\textbf{b})\Big) \quad \text{s.t}\quad \textbf{b}^T \textbf{b} \leq N_{\text{on}},
\end{equation}
where \( \mu_{\text{max}}\Big(\textbf{R}_{l,\text{eff}}(\textbf{b})^{-1}\textbf{Z}_{l,\text{eff}}(\textbf{b})\Big) \) is the largest eigenvalue of \(\textbf{R}_{l,\text{eff}}(\textbf{b})^{-1}\textbf{Z}_{l,\text{eff}}(\textbf{b})\). Although it is challenging to derive the solution of \eqref{eq:centroid_condition_opt4} in closed form, the number of possible combinations of \(\textbf{b}\) is limited due to the constraint on \( N_{\text{on}} \). 

Therefore, this work employs exhaustive searching for all possible combination of \(\textbf{b}\) under the constraint \( \textbf{b}^T\textbf{b}  \leq N_{\text{on}} \) to identify the optimal binary vector \( \textbf{b}_{\ell}^{\text{opt}} \). Once \( \textbf{b}_{\ell}^{\text{opt}} \) is determined, the corresponding optimal active current vector \( \textbf{i}^{\text{opt},}_{a,\ell} \) can be obtained by substituting it into \eqref{eq:ia_opt}. By substituting $\textbf{i}_{a,\ell}^{\text{opt}}$ and $\textbf{b}^{\text{opt}}$ into \eqref{eq:current from null space}, the full current vector can be obtained.

For the nearest neighbor condition in the GLA framework, the full current vector \( \textbf{i}_\ell \) is readily constructed using the previously obtained \( \textbf{i}_{a,\ell} \) and \( \textbf{b}_\ell \), and the condition is given by
\begin{equation}\label{eq:nearest_neighbor_rule}
    \mathcal{A}_\ell = \left\{ \textbf{h} \,\middle|\, \textbf{i}_\ell^H \textbf{H}^H\textbf{H}\textbf{i}_{\ell} > \textbf{i}_m^H \textbf{H}^H\textbf{H}\textbf{i}_{m},\ \forall m \neq \ell \right\},
\end{equation}
which implies that each channel vector \( \textbf{H} \) is assigned to the region whose corresponding current vector provides the highest received signal power.

Following the GLA, the optimal codebook can be obtained by iteratively applying the region assignment in \eqref{eq:nearest_neighbor_rule} and the centroid update in \eqref{eq:centroid_condition_opt4} and \eqref{eq:ia_opt}, starting from an initial set of binary vectors, either until convergence is achieved or for a fixed number of iterations. 
Finally, as previously mentioned, the controllable parameters at the transmitter are the voltage and the binary vector. Using the relationships given in \eqref{eq:division of ESPAR equation}, the voltage \( \textbf{v}_a \) can be computed as \(
\textbf{v}_a = \begin{bmatrix} \textbf{Z} &\textbf{Z}_{ap}^T\end{bmatrix}\textbf{i}.
\)
Based on this, the codebook \( \mathcal{B}\) can be constructed consisting of pairs of binary vectors and corresponding voltages.
We summarize the overall procedure in Algorithm~\ref{alg:proposed}.

\section{Codebook Design Based on Eigenvalue Perturbation}\label{section5}
For the codebook design described in the previous section, no closed-form solution exists for the binary vector $\textbf{b}$ that maximizes the eigenvalue, which necessitates evaluating the eigenvalues for all possible combinations of $\textbf{b}$. Exhaustively searching over all binary combinations becomes computationally intractable, even when $N_{\text{on}}$ is constrained, particularly in scenarios where the number of parasitic elements $N_p$ and codebook size $B$ are large.  
Moreover, if the current vector is expressed in the inverse form with respect to the binary vector, as in \eqref{eq:inverse_version_parasitic_current}, this process requires computing a matrix inverse for each combination, causing the overall complexity to grow exponentially.

To address this issue, we adopt the quadratic-form approximation in \eqref{eq:quadratic_of_ia} for computational efficiency, instead of using the inverse-form expression of the current vector with respect to the binary vector, given by \eqref{eq:current from null space}.
This second-order matrix $\tilde{\textbf{U}}(\textbf{b})$ not only avoids repeated inverse operations, but also enables efficient gradient-based methods due to its differentiable structure.
Leveraging this property, we propose a greedy codebook design algorithm based on eigenvalue perturbation.
\begin{algorithm}[!t]
\caption{Eigenvalue Perturbation Codebook Design}\label{alg:proposed2}
\begin{algorithmic}[1]
\STATE Construct \( \mathcal{I} = \{(\textbf{b}_1^{(0)}, \textbf{i}_{1}^{(0)}), \dots, (\textbf{b}_{2^B}^{(0)}, \textbf{i}_{2^B}^{(0)})\} \), which is the initial codebook consisting of binary and current vectors.
\REPEAT
    \STATE \( t = t +1\)
    \STATE (The Nearest Neighbor Rule) : Assign each channel vector \( \textbf{H} \) to regions \( \mathcal{A}^{(t)}_1, \mathcal{A}^{(t)}_2, ... , \mathcal{A}^{(t)}_{2^B} \) using \eqref{eq:nearest_neighbor_rule}
    \STATE (The Centroid Condition):
    \FORALL{ \( k \in \{1,\dots,N_p\} \)}
        \STATE Compute \( s_k \) using \eqref{eq:s_k2}.
    \ENDFOR
    \STATE Select $\textbf{b}_\ell^{*(t)}$ by setting ones at the indices 
corresponding to the top $N_{\mathrm{on}}$ scores among $\{s_k\}_{k=1}^{N_p}$.
    \STATE Solve GEVP with $\textbf{b}_\ell^{*(t)}$ using \eqref{eq:gen_eig} 
    \STATE Compute the active current \( \textbf{i}_{a,\ell}^{(t)} \) using \eqref{eq:ia_opt}
    \STATE Reconstruct the full current \( \textbf{i}_\ell^{(t)} \) using \eqref{eq:current from null space} subject to the power constraints in \eqref{eq:PowerConstraints}
\UNTIL{ \( \max\limits_{\ell} \left\| \textbf{i}_\ell^{(t)} - \textbf{i}_\ell^{(t-1)} \right\| < \epsilon \) \quad \text{or} \quad \( t < N_{\text{iter}} \)}
\STATE Generate the final codebook \( \mathcal{B} = \{(\textbf{v}_{a,\ell}, \textbf{b}_\ell)\} \) from \( \mathcal{I} \) by computing each $\textbf{v}_a$ using \eqref{eq:division of ESPAR equation}
\end{algorithmic}
\end{algorithm}

As can be seen from \eqref{eq:gen_eig}, this problem can be formulated as a \gls{gevp}, where both $\textbf{R}_{\ell,\text{eff}}$ and $\textbf{Z}_{\text{re},\text{eff}}$ are Hermitian matrices. Assuming that $\mu_{\max}$ is a simple eigenvalue (i.e., its algebraic multiplicity is one), the first-order eigenvalue perturbation method proposed in \cite{doi:10.1137/19M124784X} can be applied, and we have $ \partial\mu_{\text{max}} = \textbf{u}^H(\partial\textbf{R}_{l,\text{eff}} - \mu_{\text{max}}\partial\textbf{Z}_{re,\text{eff}})\textbf{u}$.
We define the initial state as $\textbf{b}_0 = [0,\ldots,0]^T$, where all parasitic antennas are turned off. For the greedy algorithm, we introduce the score $s_k$, given by
\begin{equation}\label{eq:s_k}
    s_k = \frac{\partial \mu_{\max}}{\partial b_k} = \textbf{u}_0^H \left( \frac{\partial \textbf{R}_{\ell,\text{eff}}}{\partial b_k} 
        - \mu_{\max}(\textbf{b}_0)\frac{\partial \textbf{Z}_{\text{re},\text{eff}}}{\partial b_k} \right) \textbf{u}_0,
\end{equation}
where $\mu_{\max}(\textbf{b}_0)$ and $\textbf{u}_0$ denote the maximum eigenvalue and the corresponding eigen vector under the initial state $\textbf{b}_0$, respectively. The score $s_k$ quantifies the perturbation of $\mu_{\max}(\textbf{b}_0)$ when the $k$-th parasitic antenna is switched on.

Eqn. \eqref{eq:s_k} can be expressed in terms of $\textbf{U}(\textbf{b})$ as
\begin{equation}\label{eq:s_k2}
    \begin{split}
        s_k &= \textbf{u}_0^H \Bigg(  \frac{\partial \tilde{\textbf{U}}^H(\textbf{b}_0)}{\partial b_k} \big( \textbf{R}_l - \mu_{\max}(\textbf{b}_0)\textbf{Z}_{\text{re}}\big)\tilde{\textbf{U}}(\textbf{b}_0) \\
        & \quad + \tilde{\textbf{U}}(\textbf{b}_0) \big( \textbf{R}_l - \mu_{\max}(\textbf{b}_0)\textbf{Z}_{\text{re}}\big) \frac{\partial \tilde{\textbf{U}}^H(\textbf{b}_0)}{\partial b_k}\Bigg)\textbf{u}_0 \\
        &= 2\operatorname{Re}\Bigg[ \left( \frac{\partial \tilde{\textbf{U}}(\textbf{b}_0)}{\partial b_k}\textbf{u}_0\right)^H\big( \textbf{R}_l - \mu_{\max}(\textbf{b}_0)\textbf{Z}_{\text{re}}\big)\tilde{\textbf{U}}(\textbf{b}_0)\textbf{u}_0\Bigg].
    \end{split}
\end{equation}
As defined in \eqref{eq:quadratic_of_ia}, since $\tilde{\textbf{U}}(\textbf{b})$ is a quadratic matrix with respect to $\textbf{b}$, the partial derivative of $\tilde{\textbf{U}}(\textbf{b})$ with respect to $b_k$ can be expressed as
\begin{equation}\label{eq:partial_U_of_b}
        \frac{\partial\tilde{\textbf{U}}(\textbf{b})}{\partial b_k} =
        \begin{bmatrix}
            \textbf{0}_{N_a \times 1}\\
            (\textbf{T}_k + \sum_{l=1}^{N_p}b_l\textbf{T}_{kl} + \sum_{m=1}^{N_p}b_m\textbf{T}_{mk})\textbf{Z}_{ap}
        \end{bmatrix}
\end{equation}
By substituting \eqref{eq:partial_U_of_b} into \eqref{eq:s_k2}, we can compute $s_k$. As noted earlier, $s_k$ represents the gradient of $\mu_{\max}(\textbf{b}_0)$ when the $k$-th parasitic antenna is switched on. Therefore, selecting the top $N_{\text{on}}$ indices among $\{s_k\}_{k=1}^{N_p}$ provides a low-complexity alternative to exhaustive search for determining $\textbf{b}_{\ell}^*$ such that
\begin{equation}\label{eq:select_b}
b_k=
\begin{cases}
1, & k \in \operatorname{Top}\text{-}N_{\text{on}}(\{s_k\}_{k=1}^{N_p}) \\[6pt]
0, & \text{otherwise},
\end{cases}
\end{equation}
where $\operatorname{Top}\text{-}N_{\text{on}}$ denotes the function that selects the $N_{\text{on}}$ largest elements from a given set. The initial binary vector $\textbf{b}_0$ used in this paper can be adapted based on the underlying channel statistics.
Finally, given $\textbf{b}_{\ell}^*$, we solve the \gls{gevp} in \eqref{eq:gen_eig}, compute $\textbf{i}_{a,\ell}$ as in \eqref{eq:ia_opt}, and reconstruct the full current vector $\textbf{i}_{\ell}$ using \eqref{eq:current from null space}.
Algorithm~\ref{alg:proposed2} outlines the complete procedure of the proposed method.

\begin{figure*}[t]
    \centering
    \begin{subfigure}[b]{0.32\textwidth}
        \includegraphics[width=\textwidth]{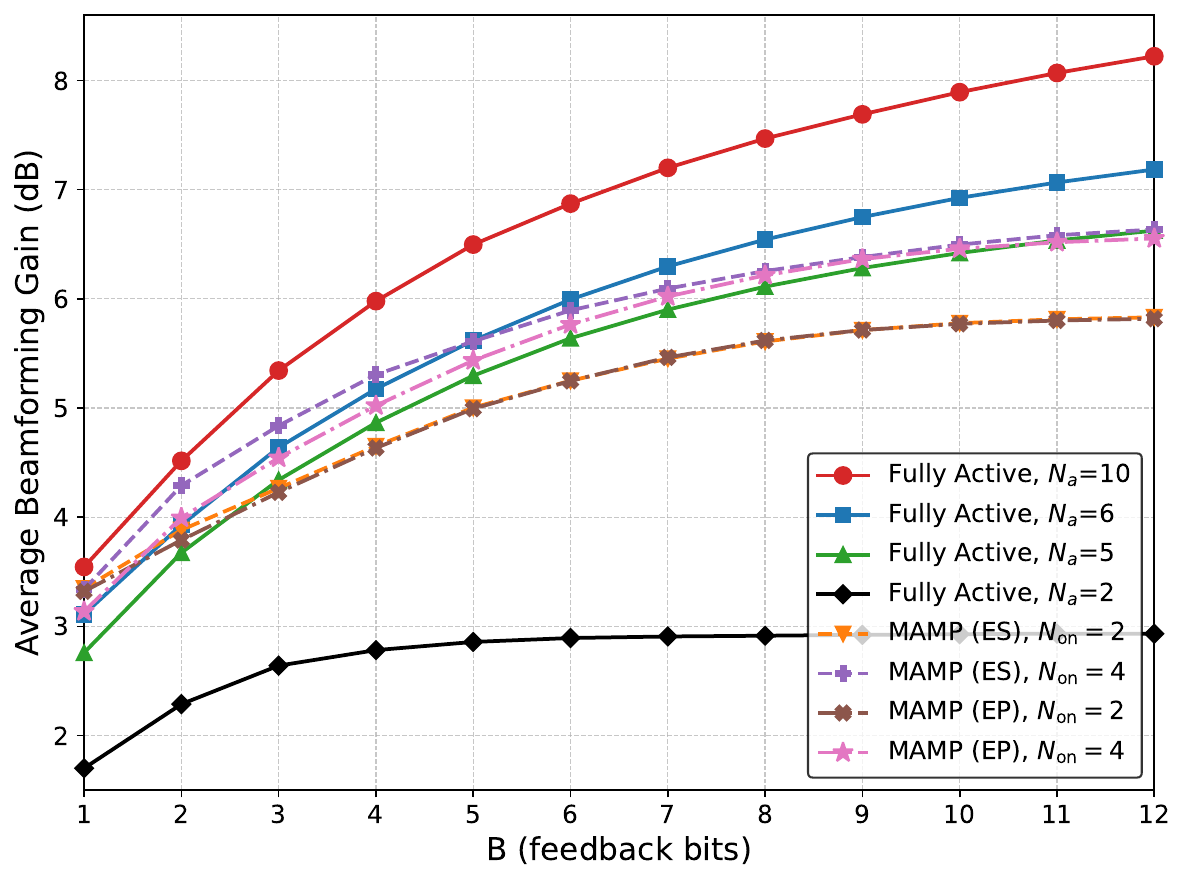}
        \subcaption{$N_a = 2$, $N_p = 8$}
        \label{fig:2A8P_iid}
    \end{subfigure}
    \hfill
    \begin{subfigure}[b]{0.32\textwidth}
        \includegraphics[width=\textwidth]{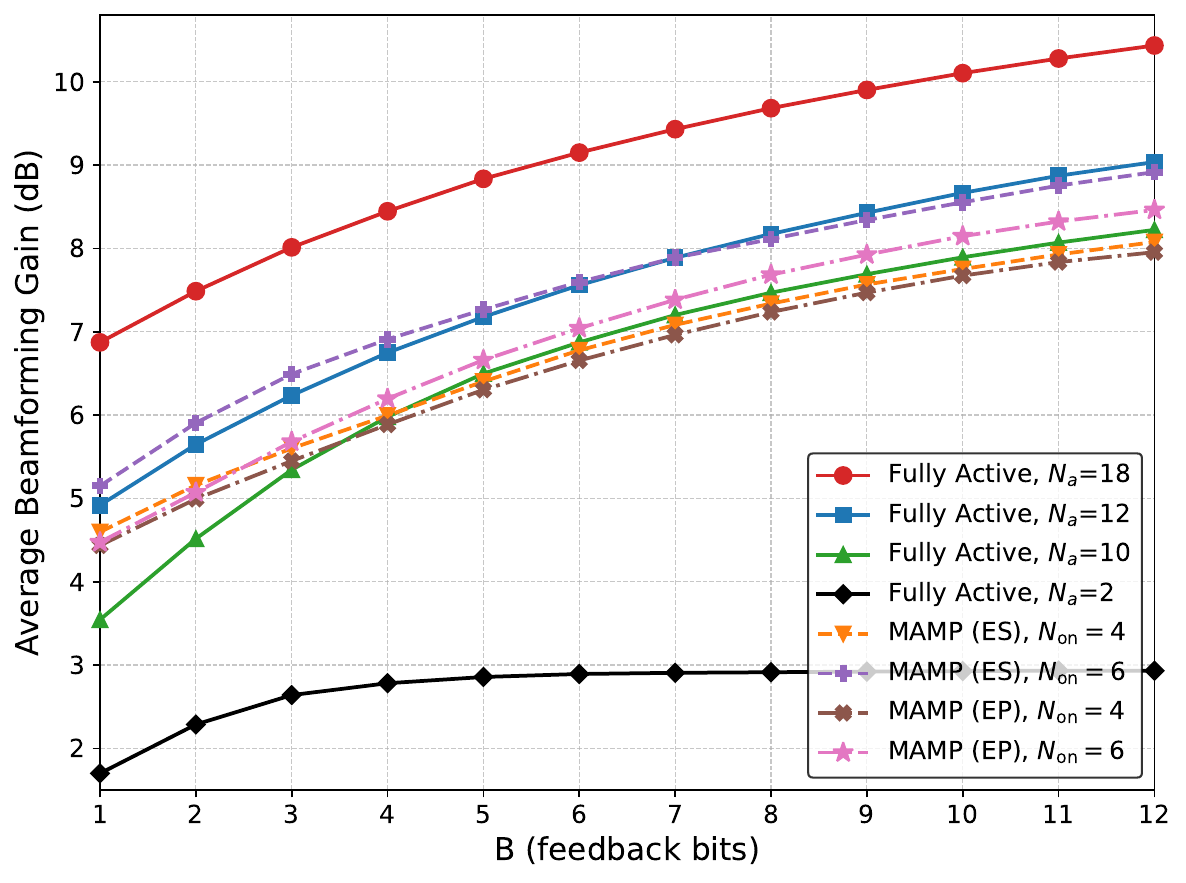}
        \subcaption{$N_a = 2$, $N_p = 16$}
        \label{fig:2A16P_iid}
    \end{subfigure}
    \hfill
    \begin{subfigure}[b]{0.32\textwidth}
        \includegraphics[width=\textwidth]{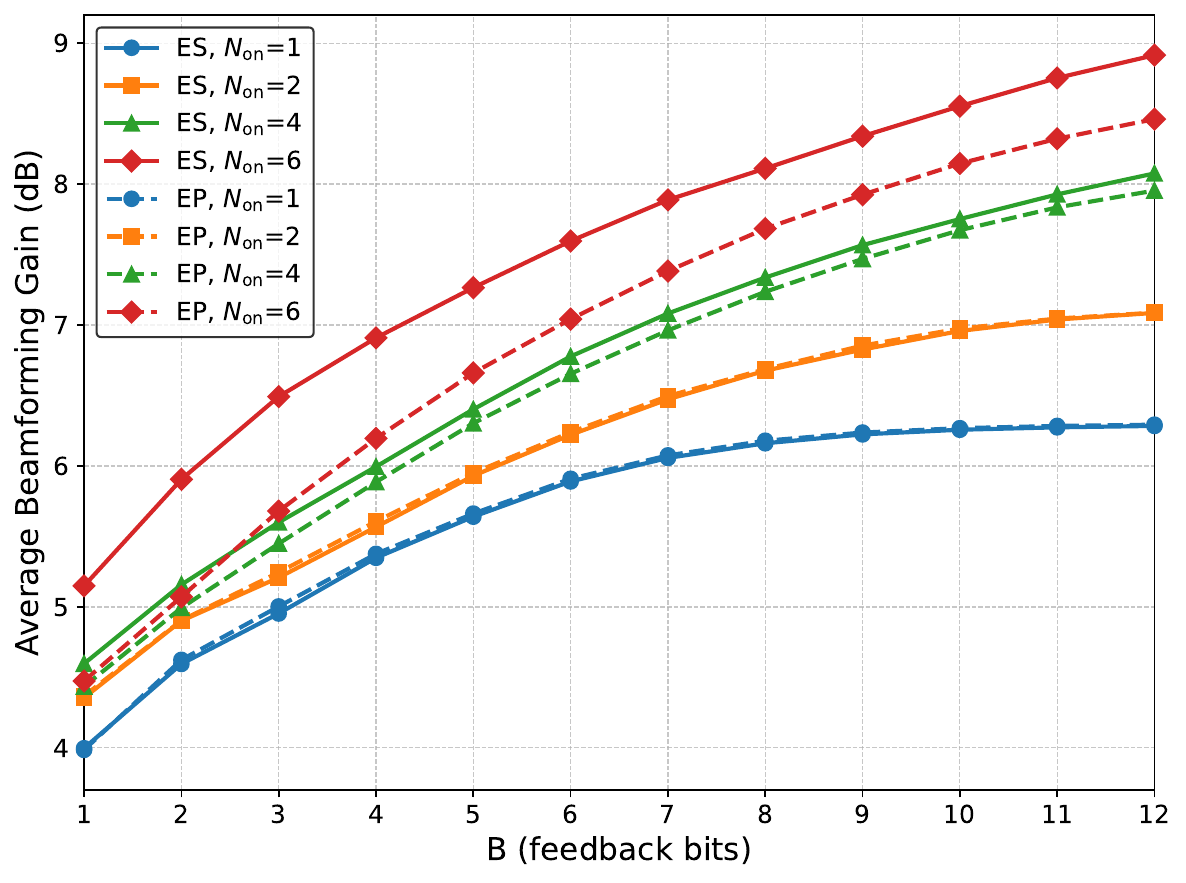}
        \subcaption{Comparison of Codebook Performances}
        \label{fig:comparison_iid}
    \end{subfigure}
    \caption{Beamforming gain versus $B$ and $N_{\text{on}}$ under i.i.d. Rayleigh channel.}
    \label{fig:iid_comparison}
    \vspace{-5mm}
\end{figure*}

\section{Simulation Results}\label{section6}
\subsection{Simulation Setup}
In this section, Beamforming performance is evaluated by varying $B$ and $N_{\text{on}}$. The transmitter operates at a center frequency of $2.4~\text{GHz}$ and employs a \gls{mamp} array composed of $N_a$ active antennas and $N_p$ parasitic antennas with two-state reactance values.
The receiver is equipped with a single active antenna so that the evaluation focuses on transmit-side beamforming. For the two-state reactance values, $x_{\text{on}}$ and $x_{\text{off}}$, we adopt $x_{\text{on}} = 80~\Omega$ and $x_{\text{off}} = -1000~\Omega$, which were determined in Section~\ref{subsec:x_on and x_off}. These values correspond to an inductance of $5.3~\text{nH}$ for $x_{\text{on}}$ and a capacitance of $66.3~\text{fF}$ for $x_{\text{off}}$ at $2.4~\mathrm{GHz}$.

The \gls{mamp} array is constructed using the optimized array parameters obtained in Section~\ref{subsec: array parameters}. It consists of $N_a$ identical subarrays, each comprising one active antenna at its center and $N_p/N_a$ parasitic antennas uniformly distributed on a circle with radius $r_1=0.3\lambda$. The spacing between adjacent active antennas is set to $r_2=0.9\lambda$.
As a reference, we consider a fully active antenna array consisting of $N_a$ active antennas, each equipped with a dedicated RF chain. To ensure a fair comparison, the distance between adjacent antennas of the fully active array is set to $r_2$, which is identical to the spacing between adjacent active antennas in the \gls{mamp} array. Linear array configurations are adopted for $N_a \leq 10$, whereas planar array configurations are considered for $N_a > 10$. The codebook for the fully active antenna array is designed using the \gls{gla} based on the centroid update in \eqref{eq:centroid_condition_opt}.

For the channel model, we assume that the receiver has perfect channel state information, and adopt a narrowband block fading channel model, where the channel remains constant until the limited feedback is updated. 
The proposed codebooks are evaluated under both i.i.d. Rayleigh fading and a realistic 3GPP channel model generated via the \gls{quadriga} simulator \cite{jaeckel2014quadriga,jaeckel2023quadriga}.

\subsection{i.i.d. Rayleigh channel model}
First, we evaluate the two proposed codebooks in terms of beamforming gain under an i.i.d. Rayleigh fading channel, which represents a highly rich scattering environment and offers a high diversity order. We set the number of iterations of the algorithm to three, which was sufficient to saturate the performance in our experiments. The channel matrix $\textbf{H} \in \mathbb{C}^{M \times N}$ is modeled as an i.i.d. circularly symmetric complex Gaussian random vector, i.e., $\textbf{H} \sim \mathcal{CN}(\textbf{0}_{M \times N}, \textbf{I}_{N})$. 
The power constraint $P_{\text{max}}$ is set such that each active antenna in the reference case with $N_a=2$ transmits with unit power.

Fig.~\ref{fig:iid_comparison} illustrates the beamforming gain with respect to $N_{\text{on}}$ and $B$ for the two proposed codebook designs. We first consider a \gls{mamp} array with $(N_a,N_p)=(2,8)$, as shown in Fig~\ref{fig:iid_comparison}(\subref{fig:2A8P_iid}). Comparing the two proposed algorithms, Algorithm~\ref{alg:proposed} and Algorithm~\ref{alg:proposed2} exhibit nearly identical performance for all values of $B$ when $N_{\text{on}}=2$. When $N_{\text{on}}=4$, Algorithm~\ref{alg:proposed} provides an additional gain of approximately $0.1$--$0.3~\text{dB}$ over Algorithm~\ref{alg:proposed} for $B \leq 6$. However, this performance gap gradually diminishes as $B$ increases, and both algorithms converge to nearly identical beamforming gains at larger codebook sizes. Compared with the benchmark, both proposed algorithms consistently outperform the fully active array with $N_a=5$ over the entire range of $B$. Furthermore, Algorithm~\ref{alg:proposed} achieves performance comparable to or better than the fully active array with $N_a=6$ for $B \leq 5$, while Algorithm~\ref{alg:proposed2} exhibits comparable performance for $B \leq 2$. These results suggest that eight switched parasitic antennas can effectively replace three to four additional active antennas, leading to significant improvements in both cost and energy efficiency.

Fig.~\ref{fig:iid_comparison}(\subref{fig:2A16P_iid}) presents the beamforming performance for the larger and more densely packed \gls{mamp} array with $(N_a,N_p)=(2,16)$. Compared with the benchmark fully active antenna arrays, Algorithm~\ref{alg:proposed} with $N_{\text{on}}=6$ achieves beamforming gains nearly identical to those of the fully active array with $N_a=12$. Likewise, Algorithm~\ref{alg:proposed2} consistently outperforms the fully active array with $N_a=10$ over the entire range of $B$. These results demonstrate that sixteen switched parasitic antennas can provide beamforming performance comparable to that of eight to ten additional active antennas.

Fig.~\ref{fig:iid_comparison}(\subref{fig:comparison_iid}) compares the beamforming performance of the two proposed algorithms as $N_{\text{on}}$ increases for the $(N_a,N_p)=(2,16)$ configuration. When $N_{\text{on}}=1$ and $2$, the two algorithms achieve nearly identical performance. For $N_{\text{on}}=4$, the performance gap remains marginal, staying within $0.12~\text{dB}$ for all values of $B$. As $N_{\text{on}}$ increases to $6$, a modest performance gap is observed due to the increased approximation error, which exceeds slightly $10\%$ as discussed in Section~\ref{subsec:x_on and x_off}. Nevertheless, Algorithm~2 still provides high beamforming gains while requiring significantly lower computational complexity than Algorithm~1. Therefore, Algorithm~2 offers a practical alternative, particularly in scenarios requiring frequent codebook updates due to user mobility or time-varying channels, where its lower complexity enables efficient adaptation.

\begin{figure*}[t]
    \centering
    \begin{subfigure}[b]{0.32\textwidth}
        \includegraphics[width=\textwidth]{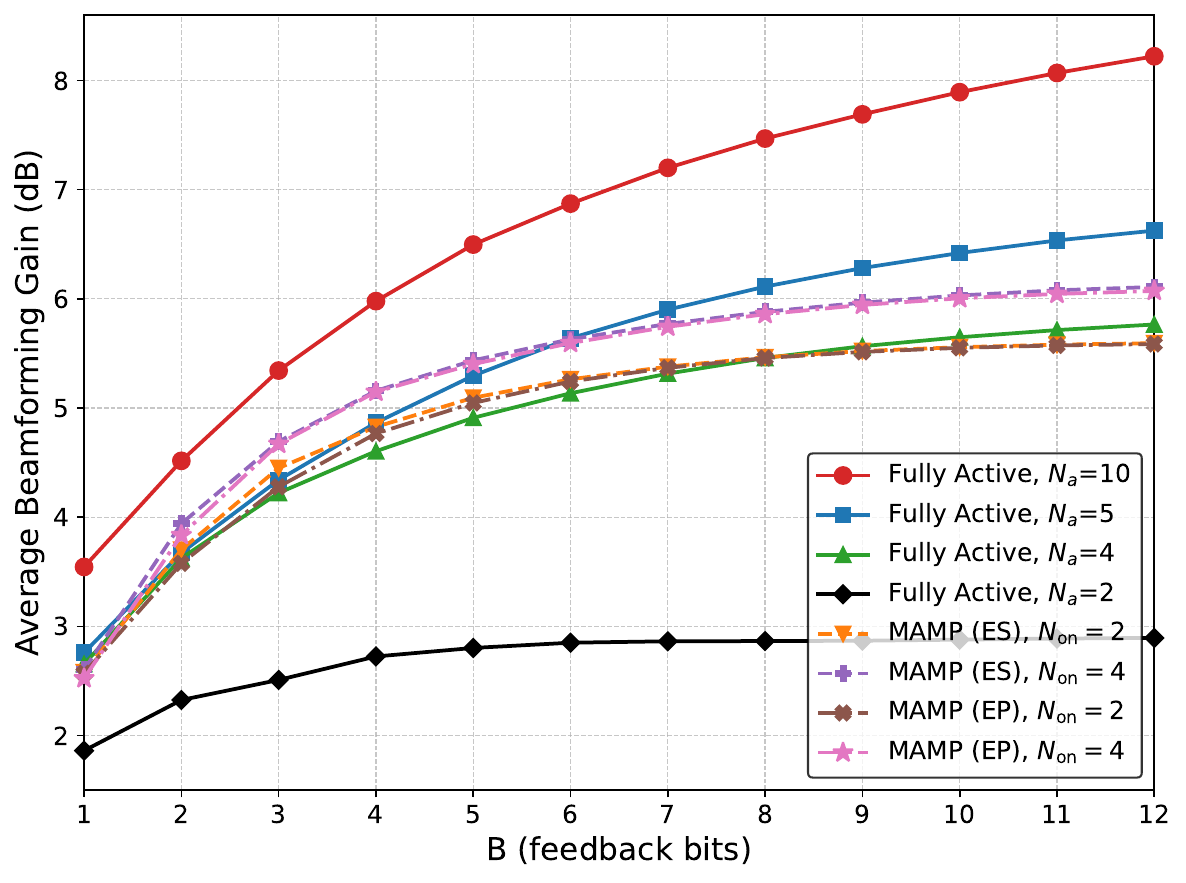}
        \subcaption{$N_a = 2$, $N_p = 8$}
        \label{fig:2A8P_UMI_NLOS}
    \end{subfigure}
    \hfill
    \begin{subfigure}[b]{0.32\textwidth}
        \includegraphics[width=\textwidth]{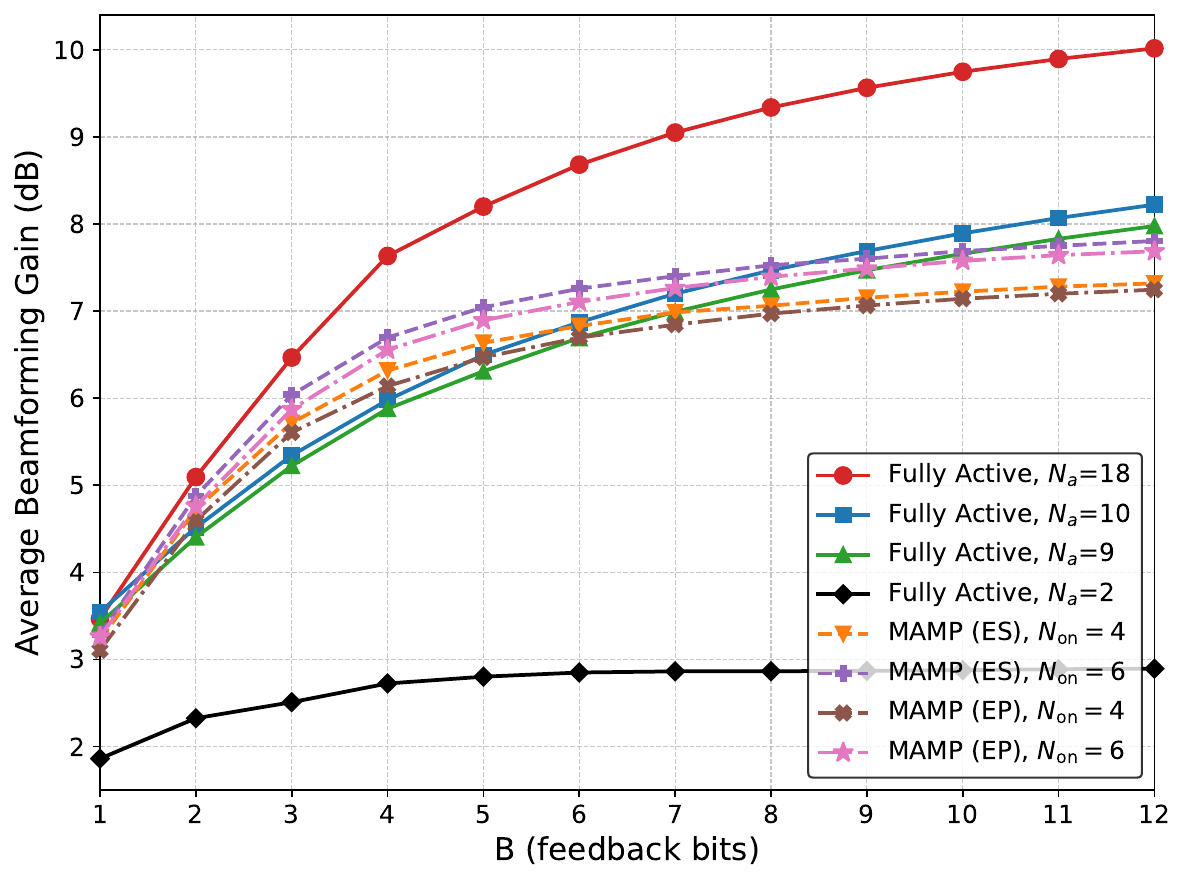}
        \subcaption{$N_a = 2$, $N_p = 16$}
        \label{fig:2A16P_UMI_NLOS}
    \end{subfigure}
    \hfill
    \begin{subfigure}[b]{0.32\textwidth}
        \includegraphics[width=\textwidth]{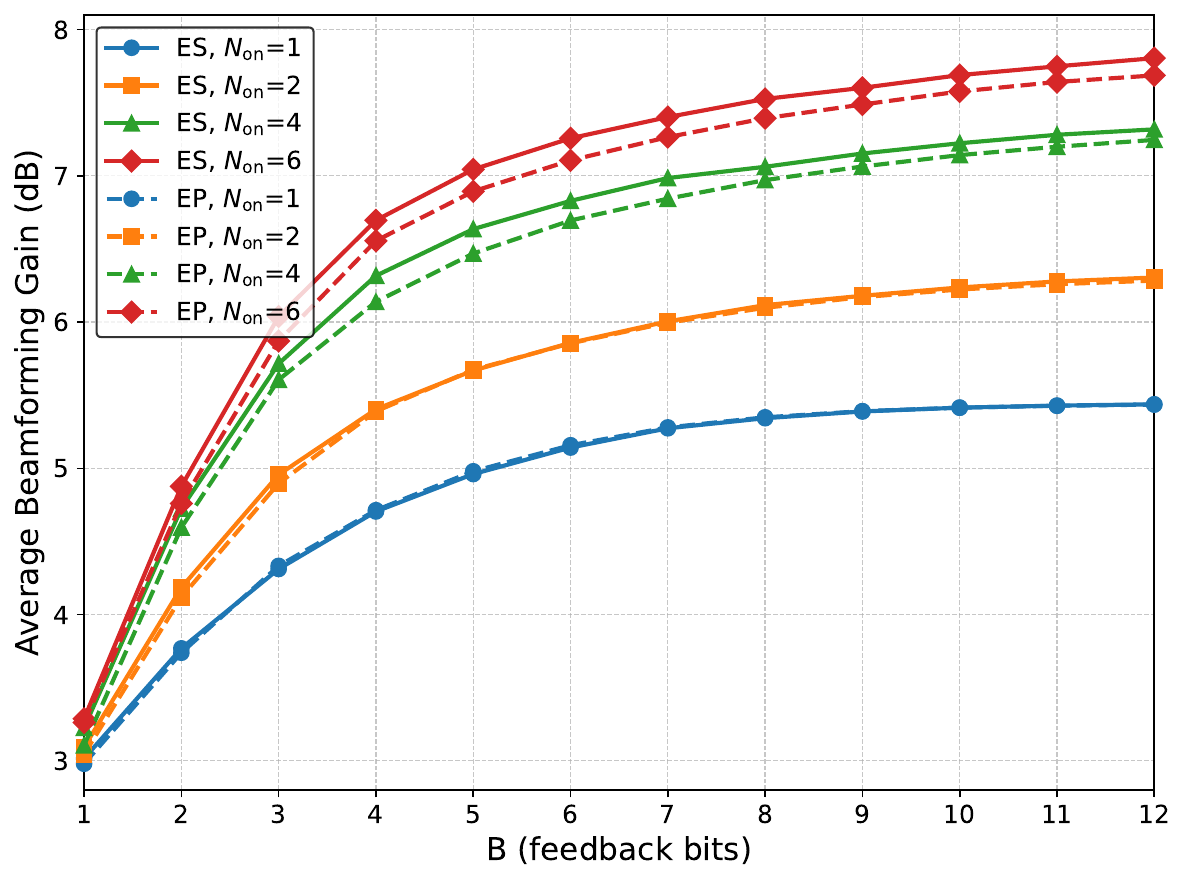}
        \subcaption{Comparison of Codebook Performances}
        \label{fig:comparison_UMI_NLOS}
    \end{subfigure}
    \caption{Beamforming gain versus $B$ and $N_{\text{on}}$ under 3GPP UMI NLOS Scenario.}
    \label{fig:UMI_NLOS_comparison}
    \vspace{-3mm}
\end{figure*}

\subsection{\gls{3gpp} \gls{umi} Scenario}
We evaluate the proposed algorithms in a realistic setting by generating \gls{umi} \gls{nlos} channel scenarios, based on the \gls{3gpp} TR 38.901 model~\cite{3gpp_tr_38_901}, using the \gls{quadriga} simulator. Considering the power constraint of the proposed \gls{mamp} array, the \gls{umi} scenario is chosen instead of the \gls{uma} scenario. Following the setup in~\cite{jaeckel2023quadriga}, the terminal and base station heights are set to 1.5~m and 10~m, respectively, which reflect typical terrestrial deployments in dense urban environments.

To isolate the small-scale fading effects, we disable large-scale fading components such as path loss and shadowing.
To introduce diversity in the angular domain, terminals are uniformly distributed along a circle of radius 100~m, covering the full azimuth range of $[0, 2\pi]$. Moreover, since we assume a block fading model, the terminal positions are fixed without time-varying trajectories or mobility. Given the geometries of the \gls{mamp} array and the reference cases, identical multi-path channel realizations are mapped to spatial channel matrices of size $\textbf{H} \in \mathbb{C}^{M \times {N}}$.
For a fair comparison with the i.i.d. Rayleigh case, the spatial channel matrices are normalized such that the average channel power of the reference model with $N_a =2$ corresponds to $3~\mathrm{dB}$. The same scaling factor is then applied to the \gls{mamp} channel model and other benchmarks.

Fig.~\ref{fig:UMI_NLOS_comparison} presents the beamforming gain of Algorithms~\ref{alg:proposed} and~\ref{alg:proposed2} under the UMi NLOS channel scenario. Fig.~\ref{fig:UMI_NLOS_comparison}(\subref{fig:2A8P_UMI_NLOS}) shows the beamforming performance for the $(N_a,N_p)=(2,8)$ configuration. The two proposed algorithms achieve nearly identical performance for both $N_{\text{on}}=2$ and $N_{\text{on}}=4$ over the entire range of $B$. Compared with the benchmark fully active arrays, both algorithms with $N_{\text{on}}=4$ consistently outperform the reference system with $N_a=4$ and achieve performance comparable to or better than the fully active array with $N_a=5$ for $B\leq6$. These results indicate that, even under a practical channel model, eight low-cost switched parasitic antennas can effectively replace two to three additional active antennas.

Fig.~\ref{fig:UMI_NLOS_comparison}(\subref{fig:2A16P_UMI_NLOS}) presents the results for the $(N_a,N_p)=(2,16)$ configuration. Both algorithms with \(N_{\text{on}} = 6\) consistently achieve performance comparable to or better than the fully active array with $N_a=9$ for all $B\leq11$, while also matching or outperforming the fully active array with $N_a=10$ for $B\leq8$. These results demonstrate that increasing the number of densely packed parasitic antennas further improves the beamforming capability, enabling the replacement of approximately seven to eight additional active antennas.

Fig.~\ref{fig:UMI_NLOS_comparison}(\subref{fig:comparison_UMI_NLOS}) compares the performance of the two proposed algorithms as $N_{\text{on}}$ increases for the $(N_a,N_p)=(2,16)$ configuration. The two algorithms exhibit virtually identical performance for $N_{\text{on}}=1$ and $2$. Even for larger values of $N_{\text{on}}$, the performance difference remains marginal, staying below $0.13~\text{dB}$ for $N_{\text{on}}=4$ and below $0.16~\text{dB}$ for $N_{\text{on}}=6$. Compared with the i.i.d. Rayleigh channel, the performance gap between the two algorithms is considerably smaller under the UMi NLOS channel model. This can be attributed to the reduced spatial degrees of freedom of channel matrices generated from a geometry-based channel model, where the impact of the approximation error becomes less significant. Therefore, under practical propagation environments, Algorithm~2 provides nearly the same beamforming performance as Algorithm~1 while offering substantially lower computational complexity, making it an attractive alternative for practical implementations.

\subsection{Computational Complexity Analysis}
In this subsection, we compare the computational complexity of the two proposed codebook design algorithms. Although both methods employ a \gls{gla} and share the same procedures for assigning channel regions and computing the sample channel covariance matrix $\textbf{R}_l$, the primary difference in complexity arises from the centroid condition, which identifies the representative solution vector for each region. Therefore, our analysis focuses on the centroid computation step.

For each region, Algorithm~\ref{alg:proposed} performs an exhaustive search over all possible combinations of $\textbf{b}$, i.e., $\sum_{i=1}^{N_{\text{on}}} \binom{N_p}{i}$ binary vectors. For each candidate $\textbf{b}$, it solves \gls{gevp}, which involves computing matrices $\textbf{U}(\textbf{b})$, $\textbf{R}_{l,\text{eff}}$, and $\textbf{Z}_{p,\text{eff}}$, followed by eigenvector decomposition. The computational cost per evaluation is approximately $O(N^3 + N_a N^2 + N_a^3)$.
Considering all binary vector combinations and regions ($2^B$ codewords), the total complexity of Algorithm~\ref{alg:proposed} in the centroid step is given by
\begin{equation}
    O\left(2^B \sum_{i=1}^{N_{\text{on}}} \binom{N_p}{i} (N^3 + N_a N^2 + N_a^3)\right).
\end{equation}

In contrast, Algorithm~\ref{alg:proposed2} first computes the scalar scores $s_k$ defined in~\eqref{eq:s_k2}, which requires approximately $O(N_a N_p + N_p^2)$ operations. Then, it selects a single binary vector based on~\eqref{eq:select_b} and solves a single \gls{gevp}, yielding a total centroid computational complexity of
\begin{equation}
    O\left(2^B (N^3 + N_a N^2 + N_a^3 + N_a N_p + N_p^2)\right).
\end{equation}

For a given $B$ under the error-free assumption, both algorithms require solving a \gls{gevp}. However, while the number of required \glspl{gevp} in Algorithm~\ref{alg:proposed} grows combinatorially with $N$ and $N_{\mathrm{on}}$, Algorithm~2 requires only one \gls{gevp} per codeword. Consequently, Algorithm~2 scales much more favorably with the antenna array size, making it well suited for offline codebook design even for large-scale antenna arrays.

\section{Conclusion}\label{section7}
We have proposed a practical and low-complexity \gls{mamp} antenna array with binary control, where each parasitic element is configured with two reactance states. Through full-wave electromagnetic simulations using \gls{hfss}, we verified that the array geometry and configuration closely match the far-field radiation patterns of the physical antenna with those of the mathematical model. In addition, we derived the induced current vector, which serves as the beamforming vector throughout the paper, as a quadratic function of the binary parasitic vector. Based on this formulation, we identified practical reactance values that minimize the modeling error while remaining realizable in RF components.

Leveraging these results, we designed two codebooks: one is an exhaustive offline codebook that provides a theoretical performance reference, and the other is a computationally efficient online codebook using eigenvalue perturbation methods. To assess the performance and explore potential use cases, we conducted simulations across various channel environments, including i.i.d. Rayleigh theoretical fading and realistic \gls{3gpp} \gls{umi} scenario generated by \gls{quadriga}. The results demonstrated that proposed \gls{mamp} with two active antennas achieves  beamforming performance comparable to that of a fully active array with up to 12 active antennas under i.i.d. Rayleigh channel and  up to 10 active antennas under practical channel model.
Furthermore, the low-complexity design based on eigenvalue perturbation provided a favorable trade-off for online update scenarios involving frequent channel updates, such as mobile or highly dynamic user environments. 


\bibliographystyle{IEEEtran}
\bibliography{references}


\end{document}